\documentclass[
aps,prd,
nofootinbib,
superscriptaddress,
tightenlines,
amsmath,
amssymb
]{revtex4}
\usepackage{bm}
\usepackage{amsmath}
\usepackage{amssymb}
\usepackage{dutchcal} 

\usepackage{slashed}
\usepackage{color, graphicx}   
\usepackage{float}      
\usepackage{rotating}   
\usepackage{caption}    
\usepackage{subcaption} 
\usepackage{hyperref}
\usepackage{xcolor}
\usepackage{array}

\usepackage[normalem]{ulem}

\DeclareMathAlphabet{\mathdutchcal}{U}{dutchcal}{m}{n}
\SetMathAlphabet{\mathdutchcal}{bold}{U}{dutchcal}{b}{n}
\DeclareMathAlphabet{\mathdutchbcal}{U}{dutchcal}{b}{n}

\newcommand{\tcal}{\mathbf{t}}

\newenvironment{Figure}
{\par\medskip\noindent\minipage{\linewidth}}
{\endminipage\par\medskip}

\definecolor{calpolypomonagreen}{rgb}{0.12, 0.3, 0.17} 
\definecolor{cadmiumgreen}{rgb}{0.0, 0.42, 0.24} 

\newcommand{\hud}{$H^{u-d}(x,t;\zeta=0,Q^2=4$GeV$^2)$ }

\newcounter{comment}

\newcommand{\bra}[1]{\left\langle #1 \right|}
\newcommand{\ket}[1]{\left| #1 \right\rangle}

\newcommand{\ord}[1]{\mathcal{O}{(#1)}}
\newcommand{\eq}[1]{Eq.~\eqref{#1}}

\allowdisplaybreaks
\begin{document}
\title{Generalized Parton Distributions from Symbolic Regression} 

%

\author{Andrew Dotson} 
\email{adots004@nmsu.edu}
\affiliation{Department of Physics, New Mexico State University, Las Cruces, NM 88003, USA}

\author{Zaki Panjsheeri}
\email{zap2nd@virginia.edu}
\affiliation{Department of Physics, University of Virginia, Charlottesville, VA 22904, USA}

\author{Anusha Reddy Singireddy} 
\affiliation{Department of Computer Science, Old Dominion University, Norfolk, VA 23529, USA}


\author{Douglas Q. Adams} 
\affiliation{Department of Physics, University of Virginia, Charlottesville, VA 22904, USA}

\author{Marija \v Cui\'c} 
\affiliation{Department of Physics, University of Virginia, Charlottesville, VA 22904, USA}

\author{Emmanuel Ortiz-Pacheco} 
\affiliation{Department of Physics and Astronomy, Michigan State University, East Lansing, MI 48824, USA}


\author{Marie Bo\"{e}r} 
\affiliation{Department of Physics, Virginia Tech, Blacksburg VA, USA}

\author{Gia-Wei Chern} 
\affiliation{Department of Physics, University of Virginia, Charlottesville, VA 22904, USA}

\author{Michael Engelhardt}
\affiliation{Department of Physics, New Mexico State University, Las Cruces, NM 88003, USA}

\author{Gary R. Goldstein} 
\affiliation{Department of Physics and Astronomy, Tufts University, Medford, Massachusetts 02155, USA}

\author{Yaohang Li} 
\affiliation{Department of Computer Science, Old Dominion University, Norfolk, VA 23529, USA}

\author{Huey-Wen Lin} 
\affiliation{Department of Physics and Astronomy, Michigan State University, East Lansing, MI 48824, USA}

\author{Simonetta Liuti} 
\email{sl4y@virginia.edu}
\affiliation{Department of Physics, University of Virginia, Charlottesville, VA 22904, USA}

\author{Matthew D. Sievert}
\affiliation{Department of Physics, New Mexico State University, Las Cruces, NM 88003, USA}


\begin{abstract}

\vspace{0.2cm}
\begin{center}
\large
    EXCLAIM Collaboration
    \normalsize
    \footnote{The first, second, third (listed in alphabetical order) and fourth authors, contributed substantially to the calculations exhibited in this paper.} 
    \footnote{The EXCLusives with AI and Machine Learning (EXCLAIM) collaboration was recently formed, with the main goal of developing a framework where ML/AI techniques are explored to address problems emerging in the phenomenology of DVES and related processes.}
\end{center}

\vspace{0.4cm}
AI/ML informed Symbolic Regression is the next stage of scientific modeling. We utilize a highly customizable symbolic regression package ``PySR" to model the $x$ and $t$ dependence of the flavor isovector combination $H_{u-d}(x,t,\xi)$ at $\xi=0$.
These PySR models were trained on GPD results provided by both Lattice QCD and phenomenological sources GGL, GK, and VGG. We demonstrate, for the first time, the consistency and systematic convergence of Symbolic Regression by quantifying the disparate models through their Taylor expansion coefficients. In addition to PySR penalizing models with higher complexity and mean-squared error, we implement schemes that test specific physics hypotheses, including force-factorized $x$ and $t$ dependence and Regge behavior in PySR GPDs. 
We show that PySR can identify factorizing GPD sources based on their response to the Force-Factorized model. Knowing the precise behavior of the GPDs, and their uncertainties in a wide range in $x$ and $t$, crucially impacts our ability to concretely and quantitatively predict hadronic spatial distributions and their derived quantities.     
\end{abstract}

\maketitle

\section{Introduction}

Recent developments in the phenomenology of QCD extend the reach of our current understanding of the internal structure of hadrons as generated from quark and gluon dynamics. In particular, it is believed that one can map out the spatial distributions of quarks and gluons in hadronic systems by studying the correlations measured in exclusive scattering processes with more than one particle detected in the final state. This class of experiments sensitive to 3D subatomic spatial structure encompasses all deeply virtual exclusive scattering (DVES) processes. 
DVES experiments are also believed to be sensitive to the quark and gluon orbital angular momentum -- the missing component in the proton spin puzzle --  and they 
are an important part of the physics program at the upcoming Electron Ion Collider (EIC), a  state-of-the-art facility specifically planned for visualizing the internal structure of the proton, as well as of both light and heavy nuclei \cite{eicyellowreport}. 

Extracting QCD matrix elements from DVES cross section and asymmetries is, understandably, a complex problem 
where the dynamic quark and gluon observables for the process, the Generalized Parton Distributions (GPDs), are multi-dimensional functions of four kinematic invariants: $x$, the parton momentum fraction of the proton, $\xi$ the fractional longitudinal momentum transfer between the initial and scattered proton, $t$, the four-momentum transfer squared proportional to the transverse component, and $Q^2$, the virtual photon four-momentum squared. GPDs enter the cross section in integrated forms known as Compton Form Factors (CFFs) (we refer the reader to Refs.\cite{Diehl:2001pm,Belitsky:2005qn,Kumericki:2016ehc} for detailed reviews on the subject). 

It is becoming increasingly clear that, even in the simplest leading order case,  the extraction of GPDs from data is intractable if the range of computational methods is not extended beyond the standard ones, {\it i.e.} including 
resources granted by artificial intelligence (AI) and machine learning (ML) based on pattern recognition, statistical analysis, and algorithmic decision making.
On the other hand, an alternative source of information on GPDs is given by lattice QCD (LQCD) calculations providing a framework for the direct computation of GPDs from fundamental QCD principles, that can be compared with experimental observables as they become available. 

Regardless of whether GPDs will be ultimately inferred from experiment and/or from the calculation of LQCD matrix elements, possibly with the aid of ML methods, the physical information they carry through their dependence in the multiple kinematic variables they depend on, will remain challenging to disentangle. It is therefore appropriate to ask the the question:

\vspace{0.2cm}
\noindent {\it Can we identify specific regularities, factorization or other symmetric behaviors in the multi-dimensional GPD data, that lead to the interpretation of specific physics content of the data?} 

\vspace{0.2cm}
\noindent Addressing the latter includes, in particular, the extraction of information on the spatial distributions of quarks and gluons inside the proton.  
Addressing the question of making progress towards the simultaneous extraction from data and interpretation of GPDs, brought us to explore new developments in {\it physics explainable AI methods} that are designed to provide a new handle to overcome the inherent, game-stopping computational hurdles of the analysis. 
Towards this goal, we 
introduce here a Symbolic Regression (SR)-based analysis of GPDs. 

SR is a relatively new AI technique that is ideal for physics applications in that it allows us to identify concise and interpretable mathematical expressions directly from data. Rooted in genetic programming, it has recently become a favored data-driven deep learning model discovery tool, in various sectors of theoretical physics. For instance, the SR package NeSymReS \cite{Biggio:2021} has recently been used to extract fragmentation functions from identified charged hadron data in semi-inclusive deep inelastic scattering at COMPASS \cite{Makke:2025zoy}, and PySR was used in Ref.~\cite{Morales-Alvarado:2024jrk} to fit the proton structure functions in the Drell-Yan process.

We apply SR to the $x$ and $t$ dependence of GPDs from LQCD calculations, at a fixed value of the other kinematic variables, $\xi$ and $Q^2$.  
SR replaces the traditional regression procedure of starting with a fixed functional form and optimizing a set of parameters, with constructing an arbitrary functional form by assembling a string of elementary operations and mutating the expression dynamically to adapt to the data.  Like neural networks and other ML algorithms, SR has the advantage of removing the systematic bias due to a frozen or truncated functional form.  But unlike other approaches, SR produces simple analytic expressions that optimize both the fit quality and the simplicity of the expression.  As such, SR has a distinct advantage with regard to \textit{interpretability}.  This results not only in a greater ease of interpretation of the range of behaviors preferred by the fit, but also in the interpretation of what types of analytic forms are more physically sound behaviors.  

The flexible analytic forms are particularly well-suited to imposing physical constraints, which often take the form of integral or differential equations.  Because of its ease of mathematical interpretation, SR can be used to greatly expedite the process of narrowing the search space of functional forms only to those which are physically reasonable.  The simplest demonstration of this is through the post-facto selection of fit functions which obey a physical constraint such as a sum rule.  But far more powerful is to directly modify the loss function used by the SR algorithm to determine the optimum functional form.  Implementing a custom loss function in SR allows us to bias the optimization process to only construct functional forms which satisfy a given physical constraint.  This allows one to reasonably explore the \textit{space of all possible fits consistent with the physical constraint}, limited by the hyperparameters of the algorithm.  Through its ability to robustly incorporate physical constraints, SR exhibits a remarkable potential for improvement in \textit{extrapolation} by balancing fit quality, complexity, and physical constraints. In turn, controlling these features is a mandatory step for evaluating the Fourier transforms that will ultimately yield the quark and gluon spatial distributions of the proton. For these reasons, SR constitutes a profound new tool for discovery of hadron structure.

Our analysis of GPDs was first motivated to find the physics in LQCD results that come straight out the Lagrangian with no physics in between. 
The initial scan of LQCD results with SR led to a first surprising outcome that the LQCD results that we had been analyzing vary independently in the $x$ and $t$ variables  (or they display a factorized dependence in $x$ and $t$): this question becomes important for visualizing the internal proton structure, since $t$ is related to the spatial distance of a given quark and gluon from the proton's relativistic center of momentum. A factorized result would imply that quark and gluon configurations carrying a different $x$ momentum fraction, all have the same radial dependence, a structure contradicting many scenarios in the phenomenology of QCD.  
In a second phase of our study, we addressed how to make this finding rigorous, including the study of the kinematic settings under which factorization happens, 
and we devised a {\it convergence criterion}, that we propose as a method that can be applied generally in SR, beyond the scope of interpreting QCD results. The convergence criterion, named Expansion Coefficient Clustering (ECC), is explained in detail in this manuscript. 
The symbolic expressions that we found are not in the format that one would directly expect from phenomenology, but they hint at new possible ways of interpreting results. All of the the stages of our analysis have led us to devise methods to enable the exploration of the physics scenarios that are under wraps in the data or LQCD results, as it was in the first attempt, only that now we added layers of stringency and self-consistency. 

This paper is organized as follows: in Section II we give an overview of GPDs and the steps involved towards their experimental observation; in Section III we introduce our approach for SR, using, specifically PySR; in Section IV we present our results for the PySR best fit of the LQCD flavor non singlet GPDs, discussing various scenarios for the $x$ and $t$ dependence and their impact on the spatial density distributions; 
finally in Section V we present our conclusions and an outlook on our future program development.

\section{Physics Framework: A Primer on Generalized Parton Distributions}
\label{sec:2}
Generalized parton distributions are matrix elements of collinear, light-like separated operators evaluated between hadronic states with different momentum and helicity \cite{Diehl:2001pm,Belitsky:2005qn}.  

\vspace{0.2cm}
\noindent $\bullet$ In the vector sector, the GPDs $H$ and $E$ parameterize the leading order in QCD matrix element as (we adopt the same notation as in Ref.\cite{Meissner:2009ww}),
\begin{eqnarray}
\label{eq:correlator}
 W^{[\gamma^+]}_{\Lambda\Lambda'} (p,p',q) & = &   \int\frac{dz^-}{4\pi} e^{i x p^+ z^-} \bra{p', \Lambda'}
    \bar{q}(-\tfrac{z}{2}) \: \gamma^+ \: 
    \mathcal{U}[-\tfrac{z}{2} \, , \, \tfrac{z}{2} ] \: q(\tfrac{z}{2}) \: \ket{p,\Lambda} \Big|_{z^+=z_T=0} \nonumber \\
   & = &\frac{\bar{u}(p',\Lambda') \gamma^+ u(p,\Lambda)}{2 P^+}   H(x, \xi, t,Q^2)    
   +   
    \frac{\bar{u}(p',\Lambda') \, i \sigma^{+\nu} \Delta_\nu \, u(p,\Lambda)}{2M (2 P^+)} \: E(x, \xi, t,Q^2)
    \: ,
\end{eqnarray}
where $p (p')$ and $\Lambda (\Lambda')$, are the incoming (outgoing) proton momenta and helicities,  respectively, $P=(p+p')/2$, $\mathcal{U}$ is a collinear gauge link (Wilson line) and we employ light-front coordinates $p^\pm \equiv (2^{-1/2}) (p^0 \pm p^3)$. There exist two independent GPDs for the two different proton helicity states (same, $H$, and flip, $E$); similar definitions can be given for GPDs in the axial-vector sector, $\tilde{H}, \tilde{E}$, as well as in the gluon and chiral odd sectors (see {\it e.g.} Ref.\cite{Diehl:2001pm} for a detailed review).  The relevant kinematic variables 
are defined as,
\begin{eqnarray}
\label{eq:kin}
Q^2  =   -(k_e-k_e')^2, \quad x_{Bj} = \frac{Q^2}{2 (pq)}, \quad t   = \Delta^2 =   -\frac{4M^2 \xi^2}{1-\xi^2} - \frac{\Delta_T^2}{1-\xi^2}   ,  \quad  \xi  =  - \frac{(\Delta q)}{2(P q)} =   \frac{x_{Bj}\Big(1+\displaystyle\frac{t}{2Q^2}\Big)}{2-x_{Bj}+ \displaystyle\frac{x_{Bj} t}{Q^2}} , \quad x = \frac{(kq)}{(Pq)}
\end{eqnarray}
where $q=k_e-k_e'$, $k$ is the struck quark momentum.

\begin{figure}[t]
\includegraphics[width=7cm]{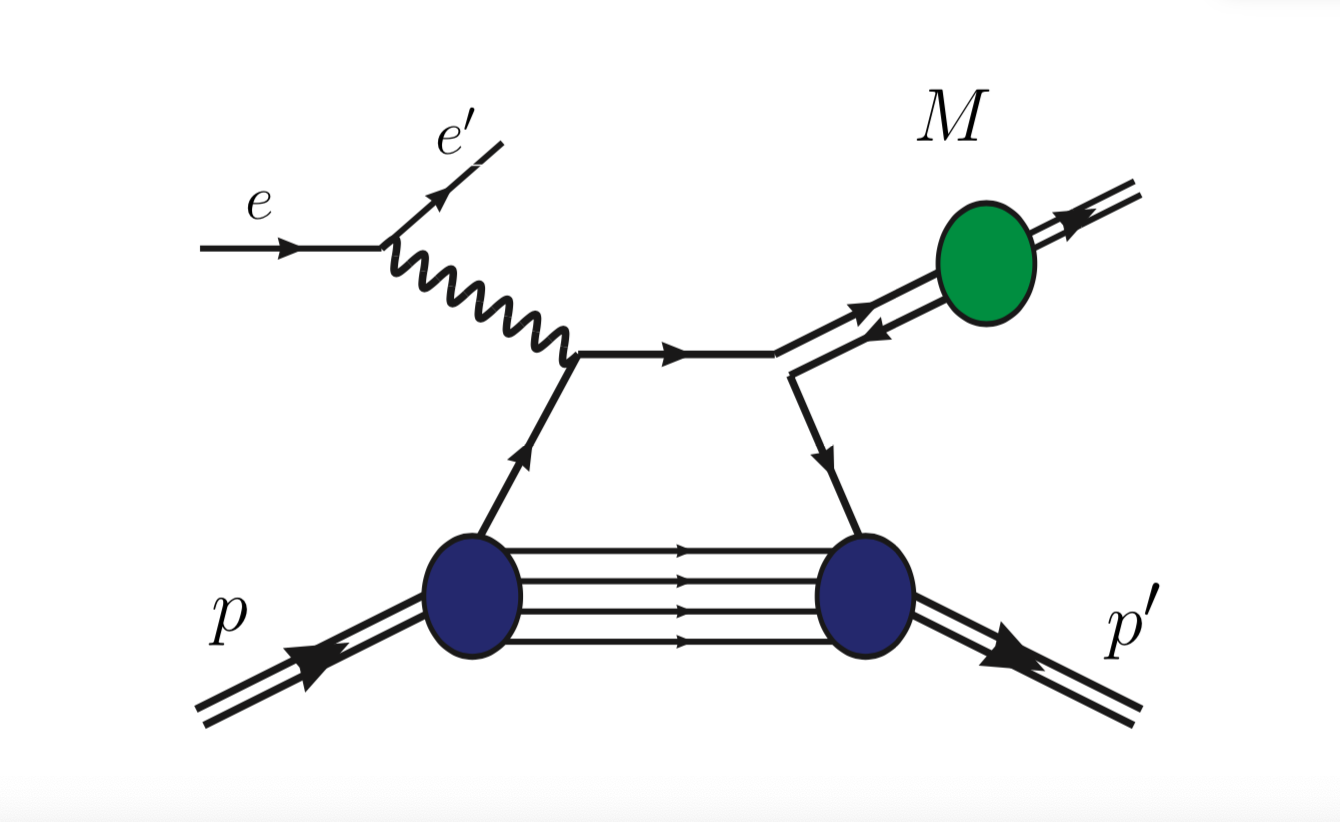}
\caption{GPD Feynman diagram at tree level, illustrating the kinematics for a deeply virtual exclusive scattering (DVES) process (Fig. \ref{fig:feynman}), $ep \rightarrow e'p' M$, where $M$ is a meson. The relevant kinematic variables are given by the hard scale, or the four momentum difference squared between the initial and final electron, $Q^2=-(k_e-k_e')^2$, the initial ($p$) and final ($p'$) proton momenta, their difference, $\Delta = p-p'$, with $\Delta^2 =t$, and the meson momentum, $q'$.}
\label{fig:feynman}
\end{figure}

\vspace{0.2cm}
\noindent $\bullet$ GPDs can be measured in experiment since they enter the parametrization of the amplitude for DVES. Within a QCD factorized picture \cite{Collins:2011zzd,Collins:2014jpa,Ji:1997nk,Ji:1998xh}, they enter the amplitude multiplied by Wilson coefficient functions calculable in perturbative QCD. GPDs define the integrated quantities over the longitudinal momentum fraction, $x$, known as Compton Form Factors (CFFs), which can be schematically written as, 
\begin{equation}
\label{eq:CFF}
    CFF(\xi,t,Q^2) = \int dx \:K(x, \xi ,Q^2) \: GPD( x, \xi,t,Q^2 )
\end{equation}
where $K$ is a perturbative QCD calculable kernel.
This factorization theorem is illustrated in the Feynman diagram for deeply virtual meson production shown in Fig.~\ref{fig:feynman}. To leading power in $Q^2$ in the factorization theorem, all DVES processes, including deeply virtual Compton scattering (DVCS) \cite{Ji:1996nm}, deeply virtual meson production (DVMP), timelike Compton scattering (TCS), are parametrized in terms of the same building blocks -- the GPDs -- with different kinematic coefficients. 

\vspace{0.2cm}
\noindent $\bullet$ The GPDs $H$ and $E$, in the vector sector, respectively define the integrands for the Dirac and Pauli form factors of the nucleon,  
\begin{eqnarray}
\label{eq:mom1}
 A_{10}(t) \equiv    \int_{-1}^1 dx  H_q(x,\xi,t,Q^2) = F_1(t) , \quad\quad  B_{10}(t) \equiv \int_{-1}^1 dx  E_q(x,\xi,t,Q^2) = F_2(t)
\end{eqnarray}
where the notation on the {\it lhs}, $A_{n0}, B_{n0}$, with $n$ integer, adopted in most of the LQCD literature, follows from the property of polynomiality, which is in turn a reflection of the QCD Operator Product Expansion \cite{Diehl:2003ny}.  
The variable $\xi$ drops out of the $n=1$ integrated forms as a consequence of Lorentz covariance.  
The $n=2,3$ Mellin moments in $x$, are defined below,
\begin{subequations}
\begin{eqnarray}
\label{eq:mom2}
 A_{20}(t,Q^2) + (2\xi)^2 C_{20}(t,Q^2) \equiv    \int_{-1}^1 dx  x \, H_q(x,\xi,t)  ,&& \quad  B_{20}(t,Q^2) - (2\xi)^2 C_{20}(t,Q^2)  \equiv \int_{-1}^1  dx \, x \,  E_q(x,\xi,t,Q^2) \\
  A_{30}(t,Q^2) + (2\xi)^2 A_{32}(t,Q^2) \equiv    \int_{-1}^1 dx  x^2 \, H_q(x,\xi,t)  , &&\quad  \!\!\! B_{30}(t,Q^2) + (2\xi)^2 B_{32}(t,Q^2)  \equiv \int_{-1}^1  dx \, x^2 \,  E_q(x,\xi,t,Q^2)
\end{eqnarray}
\end{subequations}
$A_{20}$ and $B_{20}$ give access to angular momentum, through the sum rule \cite{Ji:1996ek},
\begin{eqnarray}
\label{eq:Ji}
   \int_{-1}^1 dx  x \, \sum_q \Big(H_q(x,\xi,t,Q^2) +  E_q(x,\xi,t,Q^2) \Big) + H_g(x,\xi,t,Q^2) +  E_g(x,\xi,t,Q^2)  = J_q + J_g = \frac{1}{2} ,
\end{eqnarray}
where the subscripts $q$ and $g$ run over the different flavor quark and gluons, respectively. 

\vspace{0.2cm}
\noindent $\bullet$ Through GPDs, one can evaluate the 2D quark and gluon spatial distributions in the transverse plane with respect to the proton motion, by performing a Fourier transform in the transverse momentum transfer $\Delta_T$ (Eq.\eqref{eq:kin}, for any given $x$.  A physical interpretation is most transparent for $\xi=0$, where we can define the spatial distribution for the unpolarized quark and gluon density in the transverse plane, $\rho^{q,g}$, as \cite{Burkardt:2000za},
\begin{eqnarray}    \label{e:2D_Density}
    \rho_{q,g}(x, {\bf b}_T) = \int \frac{d^{2} {\bf \Delta}_{T}} {(2\pi)^{2}} \: e^{-i {\bf b_{T}} \cdot {\bf \Delta}_T} \: H_{q,g}(x, 0, t,Q^2)  \: .
\end{eqnarray}
By studying the combined $x$ and $t$ (or ${\bf b}_T$) dependence, one can, in principle, infer the location, or radial distribution,  of a given parton species for any given longitudinal momentum $x$, directly from data \cite{Burkardt:2000za,Burkardt:2005hp,Belitsky:2003nz,Liuti:2004hd}. Depending on the dominance of different QCD mechanisms, it has been surmised that quarks and gluons with low momentum $x$ will be further spread out in the proton volume with respect to quarks with large $x$, which, based on one-gluon exchange mechanisms, tend to concentrate in point-like configurations (see {\it e.g.} review in Ref.\cite{Dutta:2012ii}, as well as Ref.\cite{Jain:1995dd} and Ref.\cite{Brodsky:2022bum} and references therein).  

A quantitative measure of these behaviors is given by the quark RMS transverse radius $\langle {\bf b}_{T}^{2}(x )\rangle^{1/2}$ weighted by the density in \eq{e:2D_Density},
\begin{eqnarray}
\label{eq:radius}
{\bf b}_{\mathrm{RMS}}(x) =  \langle {\bf b}_{T}^{2} (x) \rangle_{q,g}^{1/2} =  \left( \frac{\displaystyle\int d^{2} {\bf b}_{T} \:   {\bf b}_T^2 \: \rho_{q,g}(x, {\bf{b}}_T) }{\displaystyle\int d^{2} {\bf b}_{T}  \: \rho_{q,g}(x, {\bf{b}}_{T}) }\right)^{1/2} 
\end{eqnarray}
It is important to notice that, if the $x$ and $t$ dependence factorizes in $H_{q,g}$, then all $x$ configurations will have the same radial distribution, or ${\bf b}_{RMS}$ will be constant with $x$. The study of partonic configurations radii  

\vspace{0.2cm}\noindent $\bullet$ Finally, the GPDs $H$ and $\widetilde{H}$, also possess connections to ordinary parton distribution functions (PDFs) in the forward limit $t = \xi = 0$,
\begin{eqnarray}
H_q(x,0,0,Q^2) \equiv f_1^q(x,Q^2) , \quad \quad \widetilde{H}_q(x,0,0,Q^2) \equiv g_1^q(x,Q^2)
\label{eq:pdf}
\end{eqnarray}
where $f_1$ and $g_1$ are the standard spin independent and spin dependent density distributions.

\vspace{0.2cm}\noindent $\bullet$ The flavor structure of GPDs is analogous to the PDFs and is summarized in Appendix \ref{subsec:symm}.

Here we notice that we may also decompose the quark flavor dependence into isovector and isoscalar components,
\begin{subequations}
\begin{align}   \label{e:H_Isovector_Defn}
    H_{u-d} &\equiv H_u - H_d \\
    H_{u+d} &\equiv H_u + H_d \\
    \end{align}
\end{subequations}
where the valence quantum numbers imply the sum rule 
\begin{align}   \label{e:Valence_SumRule_2}
    \int\limits_0^1 dx \: H_{u - d}^- (x, 0, 0, Q^2) = 1    \: .
\end{align}
(see {\it e.g.} Eq.\eqref{e:Valence_SumRules}). Analogous expressions can be written for the GPD, $E_{q,g}$. 

The various layers of convolution between the GPDs, CFFs (Eq.\eqref{eq:CFF}), and the exclusive cross section (see Ref.\cite{Kriesten:2019jep} for a detailed description),  illustrate the nested inverse problems confronting the extraction of the multi-dimensional GPDs. Characterizing spatial densities, Eq.\eqref{e:2D_Density}, requires a systematic strategy to solve this inverse problem in a challenging multi-dimensional space. 
Even doing so, it is clear that in order to extract physically meaningful quantities from data/LQCD calculations, one needs to go beyond accepting the analysis outcome as a ``black box" prediction, and understand the ``why", or an existing general rule behind it.

\vspace{0.2cm}
\section{Symbolic Regression and Training Criteria}
\label{sec:3}
Symbolic Regression (SR) is a machine learning technique which generates and adapts symbolic mathematical expressions to fit a given data set.  
%
Unlike traditional regression techniques, in which a functional form is specified {\it a priori}, thereby introducing a systematic bias, in SR the ultimate functional form is flexible.  By iteratively adapting symbolic expressions using a given set of mathematical operations, SR optimizes a loss function which takes into account not only the fit quality, or its accuracy -- minimizing the mean-square-error --  but also the complexity of the expression which ultimately provides the best ``interpretable" relation. 
By allowing for changes in the functional forms, SR eliminates much of the arbitrary subjectivity involved in fitting a data set.  In essence, the otherwise ad hoc human-led choices implicit in the standard regression procedure, are replaced by the systematic \textit{hyperparameters} and algorithmic choices of the machine learning. At the same time, differently from a standard NN model, by returning results cast in a symbolic form, SR provides insight in the behavior of data that is directly readable, and that can, therefore, conduce to uncovering the physics relations underlying the data.    
Our analysis uses the package PySR which provides the advantage of being both versatile and highly customizable (Section \ref{subsec:PySR} and   Refs.~\cite{cranmer:2023sr,SR:2023url} for a complete description of the algorithms and inputs that drive the evolution of PySR models). The SR Loss function is built with both \textit{Mean-Square Error} $(\mathrm{MSE})$  and complexity (Section \ref{subsec:complexity}).

A major hurdle in SR analyses is the {\it convergence}: with every SR run of on the same dataset one can obtain ostensibly different functional forms of the selected models which can all be considered to be equivalently accurate and of similar complexity. To extract a physics relation from data it, therefore, becomes mandatory to be able to  
discriminate one model from another. 
In Section \ref{subsubsec: BF_consistency} we introduce a criterion for the {\it Symbolic Convergence} (SC) of SR models by casting model replicas in a common basis of Taylor polynomials, and by subsequently dividing them into clusters based on the density of replicas with comparable expansion coefficients. This procedure is defined as \textit{Expansion Coefficient Clustering} (ECC). ECC further develops the notion of SC beyond what is currently implemented in the literature.

As a specific case, we address the application of SR to Lattice QCD (LQCD) results. LQCD enables calculations in the strong coupling regime of the theory -- which is otherwise intractable --  to perform simulations of the hadronic system on a discrete space-time lattice, in a framework similar to the ones used in statistical mechanics. Analogously to NNs, albeit in a different context, LQCD predictions are, therefore, similar to a ``black-box": while predicting with a quantitative error many of the properties of hadrons and their interactions, understanding and interpreting the physics mechanisms underlying these results still requires additional theoretical and computational efforts. In particular, LQCD calculations of GPDs, reviewed in Section \ref{subsec:lattice}, provide tables of values as a function of the quark and gluon longitudinal momentum fraction, $x$, similar to PDFs, as well as the invariant momentum transfer squared, $t$, between the initial and final proton. As explained in Section \ref{sec:2}, Eq.\eqref{eq:radius}, GPDs give access through Fourier transformation, to the location of quarks and gluons inside the proton with respect to a relativistic center-of-momentum. 
It is important to understand whether and how  the $x$ and $t$ dependence are correlated, for instance, whether or not  increasing the momentum $x$ of partons gives origin to a shrinking of their radial distributions, the specific quark flavor dependence of the radial distributions, and the relation between quark and gluon spatial configurations.

\subsection{PySR:  Symbolic Regression for Python}
\label{subsec:PySR}

PySR is a multi-population evolutionary algorithm for generating these symbolic expressions which applies a classic evolutionary algorithm in parallel to a collection of separate, disconnected populations of expressions (``islands'') evolving independently.  
Other codes with different implementations are available to the community, and could be equivalently adopted in our analysis, in particular, the physics inspired AI Feynman \cite{Udrescu:2019mnk}, and various other expression tree-based algorithms following Eureqa \cite{Dubcakova:2011:GPEM}. 
These SR methods, as well as several others that have been generated recently are summarized in the living review of Ref.\cite{Makke_website}.
Our choice of PySR is motivated by the ultimate goal of interfacing the SR part of our code with the other components within the broader goal of our analysis, which includes solving the inverse problem of the extraction of GPDs and spatial structure information from data (see Section \ref{sec:2}). 
By default, PySR creates and evolves $n_p = 40$ islands containing $L = 1000$ expressions each.  These expressions are strings composed of elementary mathematical operations; in this work, we use the simple binary operations
\begin{align}
\begin{aligned}
    &\mathrm{Add}[x,y]: \qquad && x + y,
    \\
    &\mathrm{Subtr}[x,y]: \qquad && x - y,
    \\
    &\mathrm{Mult}[x,y]: \qquad && x * y,
    \\
    &\mathrm{Div}[x,y]: \qquad && x \div y, \: \mathrm{and}
    \\
    &\mathrm{Pow}[x,y]: \qquad && x \wedge y.
\end{aligned}
\end{align}
PySR also allows for the implementation of unary operators such as $\mathrm{Log}[x]$, $\sin[x]$, or more complex special functions like $\Gamma[x]$, but we do not include these in the present work.  

PySR's multi-population nature allows for efficient parallelization, and it also enables ``migration'' between the islands: the incorporation of a successful expression from one island as a sub-expression within another island.  PySR allows a fixed probability $\alpha_M = 5\%$ by default to replace an expression with one of the best expressions from the other islands, as well as a fixed probability $\alpha_H = 5\%$ by default to replace an expression with one of the best expressions so far generated.  The main loop of PySR proceeds for $n_{iterations}$ number of steps, with $n_{iterations} = 40$ by default; however, as cautioned in the documentation \cite{SR:2023url}, this value is not adequate for fully converged evolution.  As such, we use $n_{iterations} = 1000$ for our main results, and we illustrate the approach to convergence through comparisons with lower values of $n_{iterations}$.  Further details about the version of PySR used in this study are given in Appendix~\ref{app:Inputs}. 

\begin{figure}[tb]
\includegraphics[width=\textwidth]{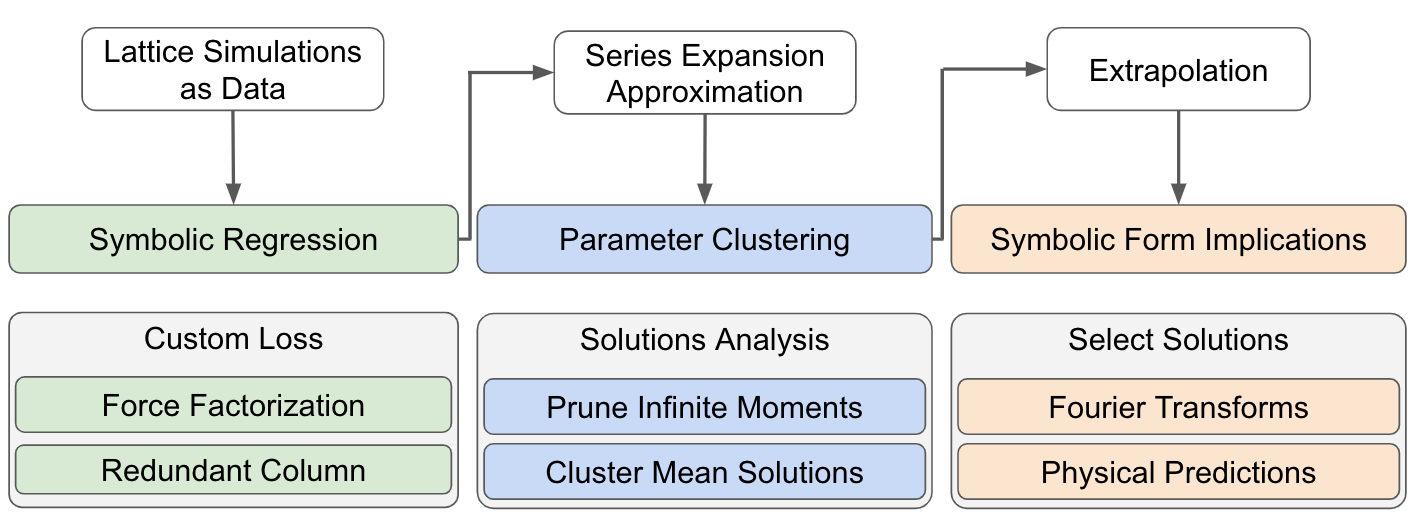}
\centering
\caption{Illustration of the procedure we used to investigate the properties of GPDs. { To gain insight into the less-explored  $x$-region of GPDs from lattice QCD,} we consider both either lattice computations or existing GPD models as "data sources" to train machine learning algorithms. { This approach extends the kinematic coverage of GPDs beyond the discrete lattice data, allowing us exploration of regions of $x$ where direct lattice calculations are challenging.} We attempt both Neural Networks (NNs) as well as Symbolic Regression (SR) as regression tools, leveraging their strengths to uncover underlying structures and potential parameterizations of GPDs in this region.}
\label{fig:flowchart}
\end{figure}

The purpose of this work is to apply the new tool PySR to generate flexible fits to the quark isovector GPD $H_{u-d} (x,t,0)$ at zero skewness $\xi=0$
computed in lattice QCD from Ref.~\cite{Lin:2020rxa}; this data is shown in Fig.~\ref{fig: PySR_HWL_train_test}.  We will also benchmark the performance of PySR and its interpretability by applying it to various well-understood microscopic models of the GPDs, as well as to compare its performance with other established ML techniques like neural networks (NN).  The various benchmarks and comparisons we perform in this work are outlined in the flowchart in Fig.~\ref{fig:flowchart}.

\subsection{Complexity and Loss}
\label{subsec:complexity}

In addition to a defined search space, search size, and training dataset, one must define a loss function (fitness function) which provides a metric for discriminating good models from bad ones and guides the overall evolution. The winning PySR model for a dataset is selected from a ``Hall of Fame" that consists of optimal expressions at each complexity. What constitutes the best model is also somewhat user-defined: one may choose to simultaneously optimize the loss function (fitness function) and complexity of candidate models, or choose against considering the complexity and simply optimize the fitness function. In the interest of attaining physically interpretable models without over-fitting, we elect to consider the complexity in the selection criteria of all PySR models explored in this work, optimizing both the loss function and complexity of these models. For this reason, a candidate PySR model can be selected over a more complex model with marginally better fitness. It has been convergence is not particularly well-defined in SR due to the nature of genetic algorithms \cite{cranmer:2023sr}, and so one expects the symbolic form of these expressions to vary, run to run. However, as discussed in Section \ref{subsubsec: BF_consistency}, we implement a convergence testing procedure on top of the PySR framework that farther develops the notion of symbolic convergence quantitatively.

When fitting an expression to a data set, it is important to give the expression enough flexibility that it can describe the general trends of the data, but not so much flexibility that \textit{over-fitting} becomes a problem.  Even in traditional regression approaches, an expression with too many free parameters and too few data points results in a fit which incorporates not just the ``true'' overall behavior of the data, but also the random fluctuations it contains as well.  In the extreme case, any set of $n$ distinct data points $(x_i, y_i)$ can always be fit exactly with an $(n-1)^\mathrm{th}$ degree polynomial.  While such an ``exact'' fit to the data is always possible, it is far from desirable; an ``optimal'' fit is one that avoids over-fitting and can successfully predict/reproduce data that were not included in determining the fit.  In addition to being less reliable at prediction, over-fit models are also more difficult to interpret.

For Symbolic Regression, in which the length and form of the expression is not fixed, over-fitting is avoided by incorporating the \textit{complexity} of the generated expression into the so-called loss function which the algorithm optimizes.  
Complexity is defined in PySR as the number of nodes in an expression tree. This definition assigns a complexity of `1' to all variables, constants, and operations. Equivalently, the complexity of an expression is then the sum of the number of occurrences of any constant, operator or variable in the expression. For example, if $\alpha,\beta$ are constants, then the common PDF expression 
\begin{align}
    x^{\alpha} (1-x)^\beta =
    \mathrm{Mult}\Big[ \mathrm{Pow}[x, \alpha] \: , \: 
    \mathrm{Pow} \big[ \mathrm{Subtr}[1 \: , \: x] \: , \: \beta \big] \Big]
\end{align}
has a complexity of 9:  one instance of ``Mult,'' two instances of ``Pow,'' one instance of ``Subtr,'' two instances of $x$, and once instance each of $1$, $\alpha$, and $\beta$.

By default, the total loss $l(M)$ of a model expression $M$ in PySR is a combination of the $\mathrm{MSE}(M)$ and the complexity $C(M)$, defined as
\begin{equation}
\label{eqn:total-loss-defn}
    l(M) \equiv l_{\text{pred}}(M)\exp{f[C(M)]}
\end{equation}
where $C(M)$ is the complexity of the model and the ``predictive loss'' $l_{pred(M)}$ is the mean-square error (MSE). The MSE is a weighted average of squared residuals for a given set of data and predictions. For a set of $N$ data points $(x_i, y_i\pm \sigma_i$ and a model $M(x)$ which evaluates to a set of predictions $(x_i, M(x_i))$ at the positions of the data, the MSE is given by
\begin{equation}
    \label{eqn:MSE}
    \text{MSE}(M) \equiv \frac{1}{N}\sum_{i=1}^N \Big( M(x_i) - y_i \Big)^2 \: 
\end{equation}

\begin{equation}
    \label{eqn:WMSE}
    \text{WMSE}(M) \equiv \frac{1}{N}\sum_{i=1}^N w_i(x_i)\Big( M(x_i) - y_i \Big)^2 \: 
\end{equation}

where, for datasets with varying point-by-point uncertainties, the weights are given by $w_i=1/\sigma^2_i$. For datasets with constant point-by-point uncertainties, we choose $w_i=1$. 
Also in Eq.\eqref{eqn:total-loss-defn}, $f[C(m)]$ is an \textit{adaptive parsimony} term called \textit{frecency} which allows for flexibility when penalizing complex expressions based on both the frequency and recency that models with that complexity have been considered by the genetic algorithm.

For phenomenological models, we set the uncertainties to 1 for the reasons described in the appendix.
$M_i^{\text{pred}}$ and $y_i^{\text{data}}$ are the model-predicted values and data being learned, respectively. In addition to  providing best-fit models for \hud that consider only the MSE and complexity, which we call `Default' selection criteria as well, we also provide PySR models that are trained on a custom loss function $l_{\text{custom}}(M)$ that, in addition to the MSE and Complexity, also penalize models that do not factorize in $x$ and $t$. This is done by introducing a non-factorization penalty $F(M)$ to the loss function (see \eq{eq:forcef}).
For expressions that factorize, $F(M)=0$ and the custom loss is simply the MSE for that factorized model. During the training, expressions that don't factorize are assigned such a large penalty that non-factorized models are forced to have lower fitness compared than the factorized ones, and thus stop being suggested by PySR. Therefore, all final losses returned by PySR are still the MSE, even in cases when a non-factorization custom loss is implemented.

\subsection{Consistency and Convergence}
\label{subsubsec: BF_consistency}
In a hypothetical world where Euler never existed, functions $f_1(x)$ and $f_2(x)$,
\begin{equation}
    \begin{split}
        &f_1(x)= e^{ix}\\
        &f_2(x)=\cos(x)+i\sin(x)
    \end{split}
\end{equation}
could be mistaken for two different functions. These two functions are equivalent, which one could deduce from their  all-orders Taylor expansions:
\begin{align}
    &e^{ix}=\sum_{n=0}^\infty \frac{(i)^n}{n!}x^n\\
    &\cos(x)=\sum_{n=0}^{\infty}\frac{(-1)^n}{(2n)!}x^{2n}\\
    &\sin(x)=\sum_{n=0}^{\infty}\frac{(-1)^n}{(2n+1)!}x^{2n+1}
\end{align}
That is, once we put these functions in a common basis (in this case, a basis of Taylor polynomials), one may make more quantitative comparisons between functions by studying any variation in the expansion coefficients. 

Because the output of PySR results are symbolic expressions, we require such a procedure in order to make quantitative statements of the ``Symbolic Convergence" of PySR models. In this section, we put model replicas in a common basis of Taylor polynomials and then cluster model replicas based on the density of replicas with comparable expansion coefficients. This procedure defines what we call \textit{Expansion Coefficient Clustering} (ECC), further developing the notion of Symbolic Convergence beyond what is currently implemented in literature. 

As a demonstration of ECC, we initialize an ensemble of 1000 unique instances of PySR and run them independently using the same initial training information. We denote these instances as PySR \textit{replicas}. These PySR replicas were trained on a particular slice of the multi-dimensional LQCD data analyzed later on this paper, 
with identical hyperparameters. In this example, the LQCD results are for a one dimensional function, $H^q(x)$, with $q=u-d$, representing a PDF calculated as defined in Eq.\eqref{eq:pdf}. 

To study the differences of these models quantitatively, we compute the $\mathcal{O}(x-c_x)^5$ Taylor expansion for each replica in the PDF limit near the point $c_x=0.555$. We choose $c_x=0.555$ because it is the midpoint of the $x$-region which contains data used to train our SR models. 
For convenience of notation let $H(x)\equiv H_{u-d}$. 
The coefficients are then defined via the Taylor expansion near $x=c_x=0.555$:
\begin{equation}
    H(x)\big|_{\text{near } x =c_x}=\sum_{n=0}^{\infty}a_n (x-c_x)^n
\end{equation}
where $a_n$ are the usual Taylor coefficients
\begin{equation}
    a_n\equiv \frac{1}{n!}H^{(n)}(x)\bigg|_{x= c_x}
\end{equation}
which involve the $n$-th derivative with respect to $x$ of $H(x)$ evaluated at $x=c_x$.
These Taylor coefficients obtained for each PySR replica can be viewed as a sample from an underlying multivariate probability distribution. 
The Taylor coefficients for a single PySR replica are also associated with an achieved MSE, and complexity. It should be remarked that the Taylor expansion basis explored here is not a unique choice: other orthogonal bases over which we can expanding the function can also be considered, and could in principle provide an optimization of our proposed ECC method, to be explored in future work.

\begin{figure}[H]
\includegraphics[width=0.9\linewidth]{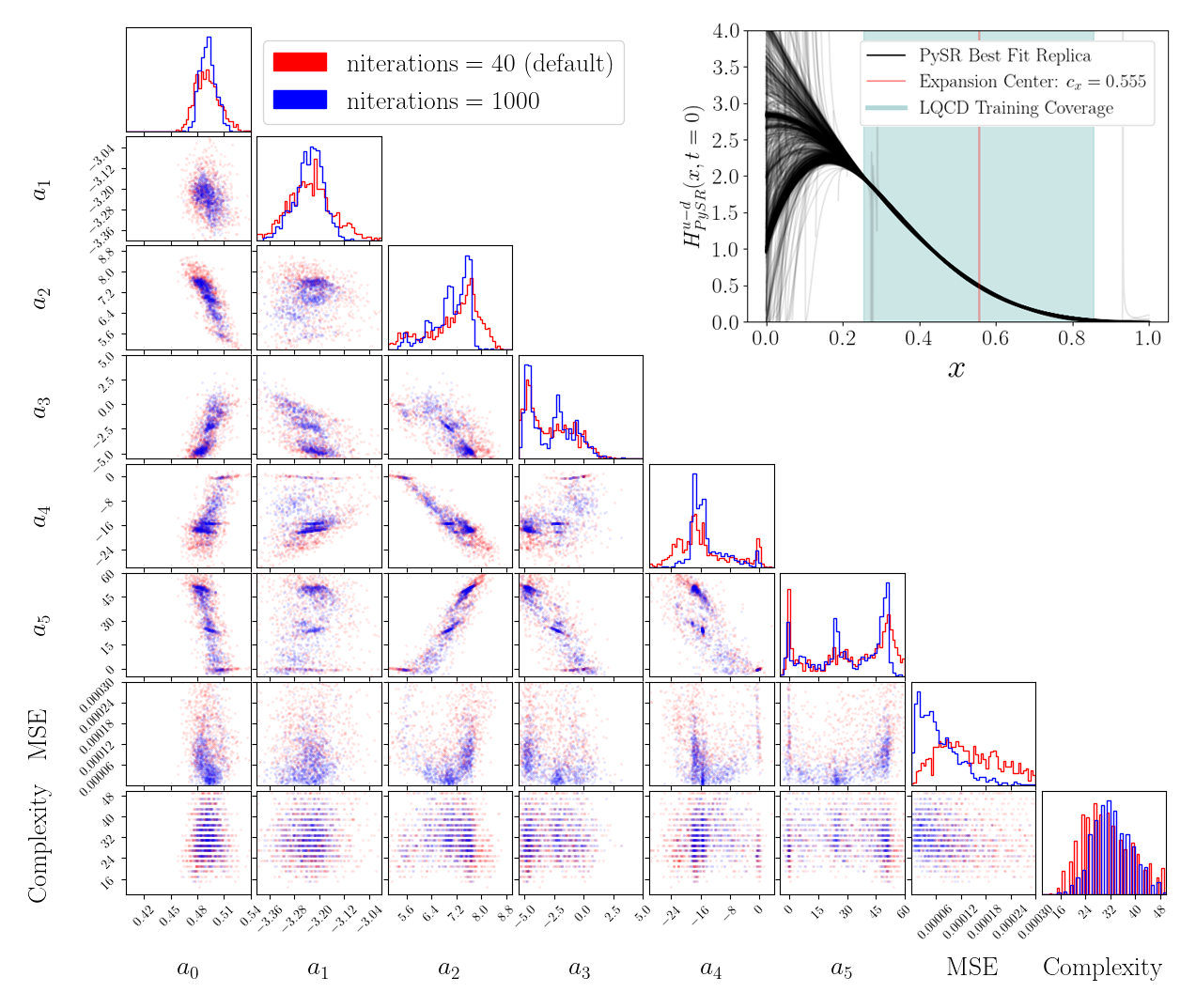}
\caption{Ensemble of 1000 unbiased fits to the lattice GPD data using PySR for $n_{iterations} = 40$ (default PySR setting, red) versus $n_{iterations} = 1000$  (blue).  The GPDs themselves for all 1000 replicas are plotted in the inset in the top-right corner (for $n_{iterations} = 1000$).  The array of smaller plots shows the distribution and correlations of Taylor coefficients $a_0 \dots a_5$,  weighted mean-square error (MSE), and complexity. Clear clusters of preferred solutions are visible by eye, and increasing $n_{iterations}$ selects those preferred solutions more precisely. 
}
\label{fig: bestFit_corner_spag_plot}
\end{figure}

A comprehensive visual picture of the coefficients, MSE values, and complexities can be seen in
Fig.~\ref{fig: bestFit_corner_spag_plot} using a corner plot \cite{corner}
for each of the 1000 PySR replicas, as well as a spaghetti plot of all 1000 un-expanded PySR models in the PDF limit.
We use a corner plot as convenient way to visualize the relevant 8 dimensional space and also provide some human intuition for correlations.

The MSEs and complexities of each PySR replica, can be systematically improved by varying initial hyper-parameters including, but not limited to, \textit{niterations} (number of generations), and \textit{maxsize} (a complexity cap). From PySR documentation, niterations = 40 is the default and chosen for speed and not performance purposes. Default hyperparameter replicas are in excellent agreement in the interpolation region, and, even in the extrapolated region where there is no training data, many models converge into potential clusters of solutions. Marginal coefficient distributions for $a_0$ and $a_1$ are clearly Gaussian, whereas higher order coefficients [$a_2$, $a_3$, $a_4$, $a_5$] exhibit multi-modal behavior.
We find such multimodal behavior is related to the behavior of these PySR replicas in the extrapolation region. 

\begin{figure}[H]
\centering
\includegraphics[width=0.5\linewidth]{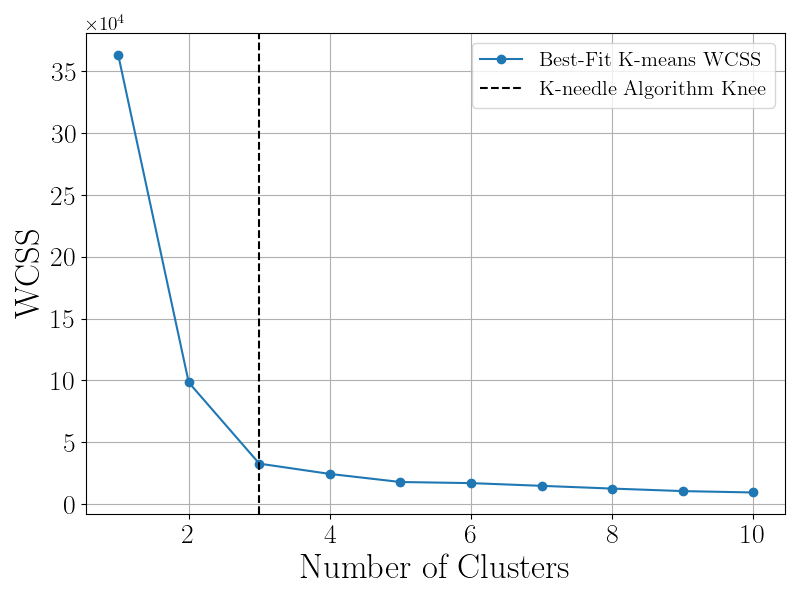}
\caption{K-means Within-Cluster Sum-of-Squares as a function of the number of clusters. Clustering is simultaneously performed on distributions $a_0-a_5$ using K-means. The location of the elbow, selected by the K-needle algorithm, suggests that these replicas are best classified into 3 clusters.
}
\label{fig: BF_elbow_plot}
\end{figure}

A natural way to characterize multi-modal behavior of samples is to understand how the samples are clustered. That is, we wish to assign clusters to the replicas which extrapolate into the different bands at low $x$ visible in the spaghetti plot in Fig. \ref{fig: bestFit_corner_spag_plot}. One may first attempt to cluster replicas based on their MSE, however, because the replicas converge to comparable MSEs regardless of their behavior in the extrapolation region, MSE is a poor parameter to control the clustering.
Instead, we perform an initial clustering analysis Taylor expansion coefficients $[a_0,...,a_5]$ using the well-accepted K-means clustering algorithm \cite{Press:2007ipz} in the 6-dimensional coefficient space. Because K-means cannot deduce the optimal number of clusters automatically, we then construct a plot of the within-cluster sum of squares (WCSS) \cite{scikit-learn-clustering} as a function of the number of clusters $k$ (Fig.~\ref{fig: BF_elbow_plot}). The WCSS is defined as follows\cite{LLoyd-alg, scikit-learn-clustering}:
\begin{equation}
    \text{WCSS}(k)= \underset{C_1,...C_k}{\text{arg min}}\sum_{j=1}^{k}\sum_{x_i\in C_j} |x_i-\mu_j|^2
\end{equation}
where $C$ is the set $k$ clusters, $\mu_j$ is the average of the samples within the $j$-th cluster ($1\leq j\leq k$), and $x_i$ is the $i$-th replica within cluster $C_j$. The selected clustering is then the ensemble of 
 $k$ clusters that minimize WCSS($k$). We use the default implementation in scikit-learn \cite{scikit-learn} for calculating the WCSS from K-means as a function of the number of clusters. The value of $k$ corresponding to the location of maximum curvature of the WCSS is referred to as the ``elbow'' or ``knee'' of the WCSS. The location of the knee was found using the Kneedle algorithm \cite{Kneedle} using the Python package kneed \cite{kneed} which uses the standard definition of the curvature $K_f(k)$ of a function $f(k)$ at any point $k$
\begin{equation}
K_f(k)=\frac{f''(k)}{\left(1+f'(k)^2\right)^{3/2}}
\end{equation}
The value of $k$ that maximizes $K_f(k)$ corresponds to the location of the Kneedle knee.
After implementing K-means and K-needle on the set of 1000 PySR Best Fit replica Taylor coefficients, we find that these replicas are best described by 3 clusters of solutions (Fig.~\ref{fig: BF_elbow_plot}).
\subsubsection{Interpolation and Extrapolation}
\label{subsubsec: BF_interp_extrap}
We can see from Fig.~\ref{fig: bestFit_corner_spag_plot} that not all replicas extrapolate to one of three clusters; There are spurious solutions that extrapolate in-between bands of overlapping models. To account for PySR models that do not fall into any cluster of solutions, we considered other clustering algorithms which do not require every single replica to lie inside a cluster \cite{scikit-learn-clustering}. We settled on using HDBSCAN to re-cluster the replicas (also using scikit-learn \cite{scikit-learn}). 
When using HDBSCAN we must set the minimum number of points per cluster, ``min\textunderscore samples". 
We found that assigning min\textunderscore samples $= 42$ is the lowest number of points that can define a cluster while still resulting in 3 clusters (+ 1 clusterless category) as an output. Taylor coefficients are just one set of parameters one could use to perform clustering; There are of course other possible approximations with other associated expansion basis functions, as well as different parameters one could consider while clustering. 

Note that clustering solely based on the MSEs of the replicas does not result in the same clusters as those achieved by clustering their Taylor coefficients. 
That is, MSE-based clusters are not mappable to the clusters found by ECC.
The ECC clusters are associated with visually consistent spaghetti lines in the extrapolation region, where as MSE clusters are not.
This can be seen as a feature (not a bug) of our novel ECC method. 
Each Taylor coefficient cluster provides a valid MSE in the data region, but a different consistent band of predictions in the extrapolation region.

\subsection{Lattice Results} 
\label{subsec:lattice}
Lattice QCD calculations provide a numerical framework for studying hadronic structure directly from quantum chromodynamics (QCD).
The present work incorporates lattice QCD data for the nucleon isovector unpolarized GPD extracted from numerical simulations performed in Ref. \cite{Lin:2020rxa} at physical quark masses, ensuring a realistic description of the nucleon. 
In particular, the calculations were carried out using a $N_f = 2+1+1$ ensemble with the highly improved staggered quark (HISQ) \cite{PhysRevD.75.054502,PhysRevD.87.054505} action for the sea quarks and the Wilson-clover action for the valence quarks. The lattice spacing is $a\approx 0.09$ fm, with a spatial (temporal) extent of 5.8 (8.6) fm,  at physical pion mass of $m_\pi \approx 135$ MeV. The nucleon matrix elements relevant for the quasi-GPDs were computed using momentum-smeared nucleon interpolating fields, and renormalization was performed in the RI/MOM scheme \cite{Lin:2020rxa,Ji:2020ect,Lin:2018pvv}. These lattice results serve as inputs for the analysis presented across this contribution. \\

\begin{figure}[H]

\includegraphics[width=10cm]{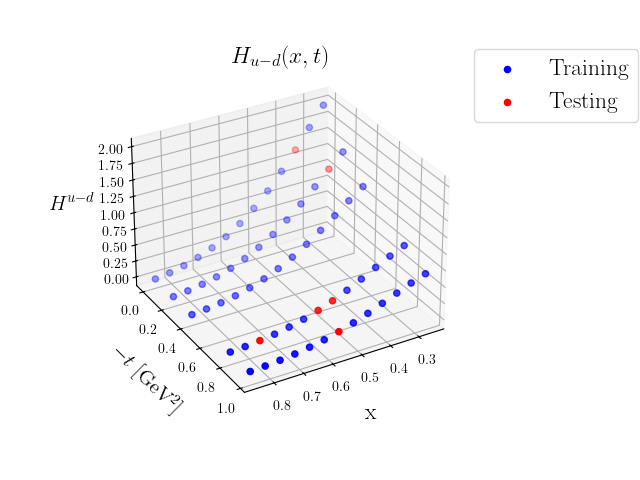}
\centering
\caption{$H_{u-d}$ results \protect\cite{Lin:2020rxa} used as input source data for training and testing SR. We can use SR to find a closed form symbolic expression for $H_{u-d}$ using the lattice result points shown in this figure. Using computational costly lattice calculations places us in a regime of "small data" where we only have about 100 points to work with.}
\label{fig: PySR_HWL_train_test}
\end{figure}

\label{subsec: training_region}
We start with a grid of $13\times 5$ lattice QCD data points in $(x,t)$, from Ref.\cite{Lin:2020rxa} 
for the nucleon isovector (flavor non-singlet) GPD, $H_{u-d}(x,t,\xi =0)$, Eq.\eqref{e:H_Isovector_Defn}, at fixed skewness parameter, $\xi =0$. The relevent renormalization scales (see \cite{Lin:2020rxa} and references therein) are
$\mu_R = 3.8$ GeV, $p_z^R= 2.2$ GeV, and $\mu^{\bar{\text{MS}}}= 3$ GeV. For brevity, we use the short-hand notation $H_{u-d}(x,t)\equiv$ $H_u$ - $H_d$ throughout. The 13 points in $x$ are linearly spaced between $x= 0.255$ and $x=0.855$, while the 5 points in $t$, fixed by the external momenta used in \cite{Lin:2020rxa}, are 
$|t|\in {0.0,0.193, 0.387,0.774,0.967}$ GeV$^2$.
We randomly remove $10\%$ of this data to reserve for testing the PySR models, while the remaining $90\%$ is used to train the models (Fig.~\ref{fig: PySR_HWL_train_test}). 

In addition to LQCD we consider phenomenological models where the functional form being learned by SR is already known. Our sources are chosen from three different models from the literature that are built according to different phenomenological criteria: a spectator model based parameterization (GGL) \cite{Goldstein:2013gra,Kriesten:2021sqc}, a double-distribution GPD (GK) \cite{Goloskokov:2013mba}, and a phenomenology based one with built-in factorized $x$ and $t$ dependence (VGG) \cite{PhysRevLett.80.5064, PhysRevD.60.094017}

For the model parametrizations we evaluate the sources on the grid of points in $(x,t)$ for which we have lattice data available and apply the same partitioning into training/testing data.

\subsection{Selection Criteria and Training Regions}
\label{subsec:selection}
In addition to training PySR on a variety of sources for the flavor non-singlet GPD combination, $H_{u-d}(x,t)$, we also provide PySR models with various functional features in mind, which we refer to as \textit{selection criteria}. 
All models consider at least a default selection criteria, which optimizes the mean-squared error and complexity of the models. The {selection criteria} we consider are:  
(1) Best Fit
(2) Force-Factorized and
(3) Semi-Reggeized.

\subsubsection{Best-Fit}
This selection criterion corresponds to a loss function where the objective is the training MSE of the PySR model, optimized against the complexity. There is no biasing towards specific functional forms or satisfying constraints, aside from the model building blocks as earlier defined. 

\subsubsection{Force-Factorized}
\label{sec: pysr_FF}
Force-Factorized (FF) corresponds to a modification of the custom loss function to assign a penalty in the training to expressions that do not factorize in $x$ and $t$. 
A function of two variables $x,t$ factorizes if it can be written as,
\begin{equation}
\label{eq:factor}
    H_{u-d}(x,t)= f_1(x)f_2(t)
\end{equation}
then forming ratios at fixed $t_i,t_j$ gives
\begin{equation}
    \begin{split}
        \frac{H_{u-d}(x,t)}{H_{u-d}(x,t)}&=\frac{f_1(x)f_2(t)}{f_1(x)f_2(t)}=\frac{f_2(t)}{f_2(t)}
    \end{split}
\end{equation}
where the ratio depends on the choices of $t_i$ and $t_j$, but is constant in $x$. We define the fixed-t ratio 
\begin{equation}
    R(x,|t|)\equiv\frac{H_{u-d}(x,|t|=0)}{H_{u-d}(x,|t|)}
\end{equation}
where the $t_i$ are values of $t\neq 0$ defined by the lattice grid in Section \ref{subsec: training_region}. For factorized forms for $H_{u-d}$, $R_i$ should be constant for all $x$.

In the FFcustom loss, a factorization penalty is assigned through an additive term  if the model for $H_{u-d}$ does not have a factorized form \ref{eq:factor}.
for two expressions $f_1$ and $f_2$ joined at a node through either multiplication or division. 
In other words, in addition to minimizing model MSEs, a penalty throughout the training if candidate models for $H_{u-d}(x,t)$ do not factorize in $x$ and $t$.
A custom loss is a fitness function that allows one to nudge PySR models towards functional forms that optimize user-defined objectives. Here, similar to \cite{Cranmer:FF-LOSS}, we  define the FFcustom loss which is implemented at each iteration of the training:
\begin{equation}
\label{eq:forcef}
    l_{custom}(M) = MSE(M) +F(M), \quad F(M)  \begin{cases}
        =0, & \text{if }H_{u-d}(x,t) = f_1(x)f_2(t)\\
        += \gamma& \text{if root degree != 2}\\
        += \gamma & \text{if root operator != $\times$ or $\div$}\\
        +=\gamma^2& \text{if not of form $f_1(x)f_2(t)$}
    \end{cases}
\end{equation}
where $f_1$ and $f_2$ are functions of either $x$ or $t$ and $\gamma = 10$ is a scale much greater than the typical MSE. The degree of the root node specifies whether the expressions are related via binary operation (degree = 2), or unary operation (degree = 1). It is impossible for an expression to factorize without two expressions at the root node being related via binary operation, and so a small penalty is assigned if that is not true for that current iteration in order to begin nudging the PySR model evolution towards the desired factorized form. For the same reasoning, the binary operator relating the two expressions at the root node must be either multiplication ($\times$) or division ($\div$), and so another small penalty is assigned if this is not satisfied. If those two criteria are satisfied, the algorithm then checks if the two expressions joined at the root node via multiplication or division are univariate and that neither expression contain the variable contained in the other expression, else a larger penalty is assigned. Splitting up the penalty into 3 smaller penalties allows the models to evolve towards increasingly fit functional forms without being heavily penalized for not immediately having the desired factorized form. 

Using the custom loss function in Eq.\eqref{eq:forcef}, all PySR replicas trained on the sources in this work were successfully factorized in $x$ and $t$. Note that this is not a statement that the training source data itself factorizes, but is rather a constraint on the space of models that the algorithm is permitted/encouraged to explore.

\subsubsection{Semi-Reggeized}
To motivate the last selection criteria considered in this work, we recall that the sum rule for up and down valance quarks PDFs:
\begin{equation}
    \int dx \left(u_v(x,Q^2)-d_v(x,Q^2)\right)= 1
\end{equation}
implies that the isovector combination $H_{u-d}(x,t)$ approaches a finite value as $x\rightarrow 0$. This is related to the scaling predicted by Regge theory \cite{Regge-theory}, in which hadronic cross sections scale as $\sigma \sim s^{2(j-1)}$, where $s$ is the center-of-mass energy (squared) and $j$ is the effective angular momentum of the $t$-channel mediator \cite{Regge-theory,Kovchegov:2012mbw}.  Collinear factorization relates cross sections like DIS to PDFs with $x \approx Q^2 / s$, leading to the general expectation that $f(x) \sim x^\alpha$ at small $x \ll 1$. While the gluon and sea quark distributions grow at small $x$ ($\alpha < 0$), the valence quark distributions fall at small $x$ ($\alpha > 0$), and the isovector component even more so.  
The behavior of this GPD near the endpoint $x \rightarrow 1$ is similarly governed by the probability to find the other two valence quarks at $x \rightarrow 0$ in the proton wave function.  This is accordingly governed by the Brodsky-Farrar quark counting rules \cite{PhysRevLett.31.1153} which predict the vanishing of PDF at large $x \rightarrow 1$ as $f(x) \sim (1-x)^\beta$, with the exponent $\beta > 0$ depending on the number of valence spectators.  In essence, the probability that a struck valence quark carries 100\% of the protons momentum should vanish at a predictable rate. 
We therefore also include a \textit{Redundant Column Approach} (RCA), which nudges PySR evolution towards models that have the form 
\begin{equation}
\label{eq:reggeized_1}
    H_{u-d}(x,t)=x^\alpha(1-x)^\beta P(x) \, g(t)
\end{equation}
where the $x$-dependent form inspired by both Regge theory \cite{Regge-theory} and the Brodsky-Farrar quark counting rules \cite{PhysRevLett.31.1153} is commonly used in parton distribution function analyses (see {\it e.g.} Ref.\cite{NNPDF:2021njg} and references therein).  Here also $\alpha$ and $\beta$ are parameters, $g(t)$ describes the unconstrained $t$-dependence of $H_{u-d}$, and $P(x)$ describes any residual x-dependence. This parameterization should give a corresponding PDF $f_{u-d}(x)$ that is finite at the endpoint regions $x\in{0,1}$.
In general, the exponents such as $\alpha =  \alpha(t)$ can also depend on $t$ in a nontrivial way (the ``Regge trajectory'').  For simplicity, in this work we consider the case of constant ($t$-independent) quantities $\alpha$ and $\beta$ and refer to such a fit as being \textit{Semi-Reggeized}.
%
%
In RCA we introduce the redundant variable $y\equiv 1-x$ so that $H_{u-d}=H_{u-d}(x,y,t)$. PySR models are then selected post-facto that have the functional form
\begin{equation}
\label{eq:reggeized_2}
    H_{u-d}(x,y,t) = x^\alpha y^\beta P(x)g(t)
\end{equation}
Taking the log of both sides gives\footnote{The transformation to log format increases the probability of the regressor to find the desired form as a linear combination of terms due to the fact that PySR-selected models that utilize the binary operator \textit{pow} manifestly break the semi-Reggeized functional form. Therefore, dropping that operator from consideration increases the probability of finding a suitably Reggeized form compared to performing SR on the original source data.
}
\begin{equation}
\begin{split}
 \ln H_{u-d}(x,y,t)&=\ln\bigg(x^\alpha y^\beta P(x)g(t)\bigg) \\
 &=\alpha\ln x+\beta\ln y+\ln P(x)+\ln g(t)
 \end{split}
\end{equation} 
Numerical results for the three different frameworks are presented and discussed in Section \ref{sec:Phen_benchmarks}.

\section{Interpreting the trends of Lattice QCD results}
\label{sec:Phen_benchmarks}
We now present results illustrating, in particular, how the use of SR for GPD analyses enables the extraction of physics information, from the spatial radii of quark and gluon distributions, to the orbital angular momentum content of the nucleon, as obtained directly from data.   

\subsection{Fit results}
\label{subsec:results}
We first show numerical results of the fits for the various selection criteria described in Section \ref{sec:3}, in particular for BF, FF, and we analyze the structures appearing in results as distinct clusters of curves. Since we do not impose a normalization criterion, namely the functional forms are not required to integrate to the same quantity, we illustrate and discuss the distribution around the nominal integrate value obtained from the form factor, $A_{10}$ in Eq.\eqref{eq:mom1}.

\subsubsection{Best Fit}
We expand the PDF limit of $10^3$ Best-Fit PySR replicas for $H_{u-d}(x,t)$ near $x=c_x=0.555$ to $\ord{x-c_x}^5$ (Fig.~\ref{fig: bestFit_corner_spag_plot}) 
Analyzing this case we encountered the issue that not all PySR replicas appear as free of pathologies: poles are visible in both the interpolation region (marked in gray in the Figure) and in the extrapolation region. 
Some PySR curves introduce poles either in-between training data points or in the extrapolation region in order for the model to fit the source data. 
%
These poles can be attributed to the model building blocks in this paper consisting of the binary operators $(+,-,\times,\div,$ \^{}). From these operators, one may construct polynomials, ratios of polynomials, powers of polynomials, and any combination. 

\begin{figure}[ht]
    \centering
    \begin{subfigure}[b]{0.45\textwidth}
        \centering
\includegraphics[width=\textwidth]{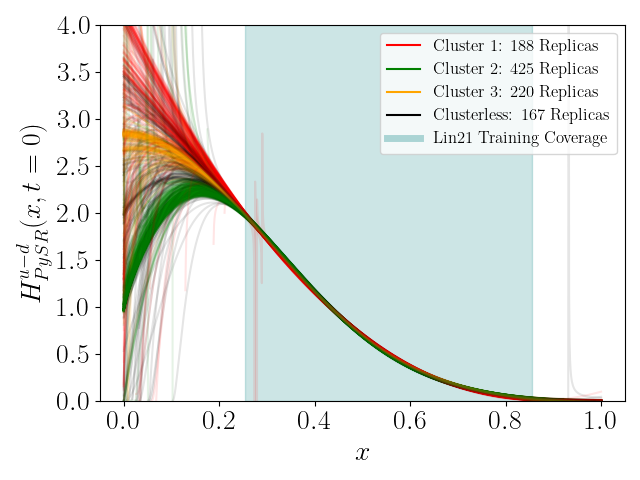}
        \caption{HDBSCAN clustering of 1000 Best-fit PySR replicas into 3 clusters + 1 clusterless category based on location of K-needle knee.}
\label{fig: BF_hwl_kneedle_spag}
    \end{subfigure}
    \hfill
    \begin{subfigure}[b]{0.45\textwidth}
        \centering
\includegraphics[width=\textwidth]{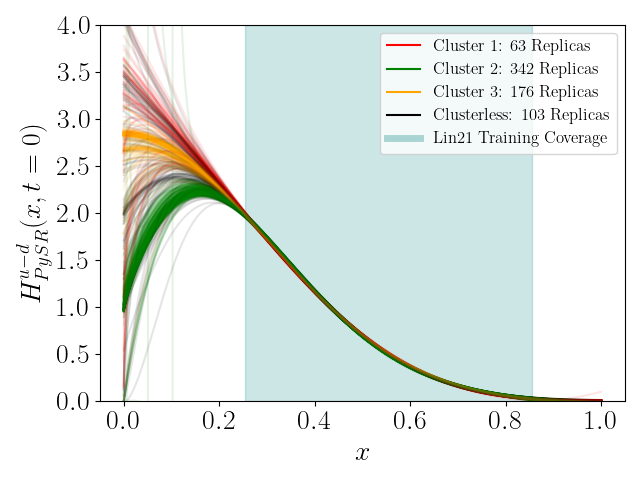}
        \caption{Remaining 684 Best-Fit PySR replicas (clustered with HDBSCAN and K-needle) after imposing Finite-Moment-Filter.}
\label{fig:BF_hwl_good_kneedle_spag_b}
    \end{subfigure}

    \caption{Cluster Color-Coded spaghetti plot of Best-Fit PySR replicas before (left) and after (right) selecting onto replicas with finite $A_{10}(t=0)$. 1000 PySR replicas trained on LQCD source data \cite{Lin:2020rxa} were generated and their PDF limit (t=0) was expanded near $x=0.555$ to $\ord{x-0.555}^5$. K-means clustering was run on the distributions of expansion coefficients simultaneously, with 3 clusters of replicas found based on the location of the K-needle knee. These coefficients were then re-clustered using HDBSCAN for the additional ``clusterless" category at fixed N=3 clusters. 
    }
    \label{fig:BF_cluster_spag}
\end{figure}

Our procedure to deal with poles solutions is to filter them out by checking whether or not the PySR model has a well-defined integral $\int_0^1 dx H_{u-d, PySR}(x,t=0)\equiv A_{10}(t=0)$, using numerical integration. 
PySR replicas with finite $A_{10}(t=0)$ define what we refer to as \textit{good replicas} throughout due to the value of this moment being connected to the sum rule in Eq.\eqref{eq:mom1}. 
The latter represents a physical constraint, in that we identify $A_{10}(t=0)$ as the quarks $u-d$ charge given by the value of the corresponding flavor dependent Dirac form factor, Eq.\eqref{eq:mom1}, measured in Refs.\cite{Cates:2011pz,Qattan:2012zf}. It is also a self-consistent check from LQCD since this quantity can be calculated in an independent framework from Ref.\cite{Lin:2021brq} (see {\it e.g.} Refs.\cite{Alexandrou:2019ali,Alexandrou:2018sjm,Bhattacharya_2023}). 
%
Out of the initial set of 1000 Best Fit replicas, 684 survive the finite-moment filter and were classified as good replicas. 
Cluster color-coded spaghetti plots of the BF replicas before and after imposing the finite-moment filter are shown in Fig.~\ref{fig: BF_hwl_kneedle_spag}.  The finite-moment filter can also be implemented into the training at the level of the custom loss function. 
We see from Fig.~\ref{fig:BF_hwl_good_kneedle_spag_b} that 342/684 good BF replicas are categorized into cluster 2, 176/684 into cluster 3, and only 63/684 into cluster 1. Many of the cluster 1 replicas diverge near $x=0$, forcing them to be discarded after imposing the finite-moment filter. Cluster 2 and Cluster 3 replicas tend towards a finite value for $H_{u-d}(x,;t=0)$ near $x=0$, resulting in a considerably higher ratio of surviving replicas for those clusters. 

\begin{figure}[H]
\centering
\includegraphics[width=0.6\linewidth]{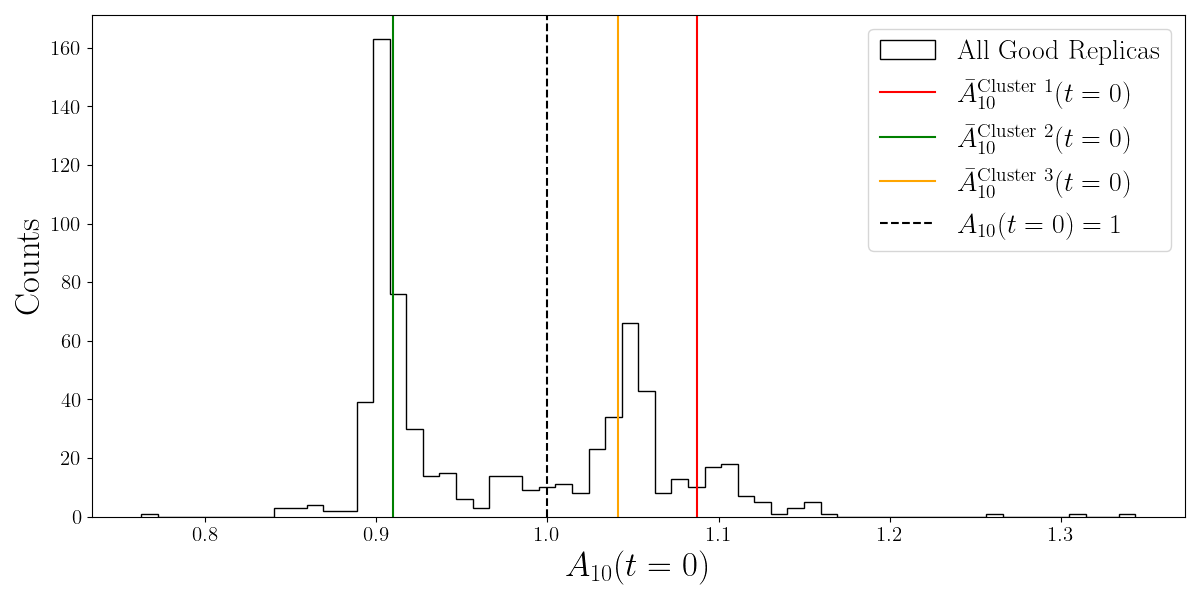} 

\caption{Distribution of finite-moment PySR Best-Fit replicas $A_{10}(t=0)$. 
Each replica is an independent run of PySR upon the same LQCD computations.
The moments of the SR replicas are centered near 1 on average, which appears to have arisen without being enforced by construction. 
Each cluster also has a mean moment which is color-coded based on HDBSCAN clustering.
}
\label{fig: BF_1st_moment}
\end{figure}

Finally, although the finite-moment filter requires that good replicas have finite $A_{10}(t=0)$, we placed no restriction that the good replicas satisfy exactly the sum rule $A_{10}(t=0)=1$. 
Interestingly, as shown in Fig.~\ref{fig: BF_1st_moment}, most of these surviving replicas still satisfy $A_{10}(t=0)=1$ to within $10\%$. 
Because the sum rule involves integrating over the full range in $x$, rather than just the training region $.255\leq x\leq 0.855$, we view this as a success in the extrapolation capability of PySR-selected models.
The physical constraint provided by integration to the elastic form factors, along with additional ones, including for instance positivity of the solutions, can be inserted directly in the custom loss function as a means of gaining additional predictive power in the extrapolation region.

\subsubsection{Forced Factorization}
\label{subsubsec: FF_consistency}
\begin{figure}[H]
\centering
\includegraphics[width=0.7\linewidth]{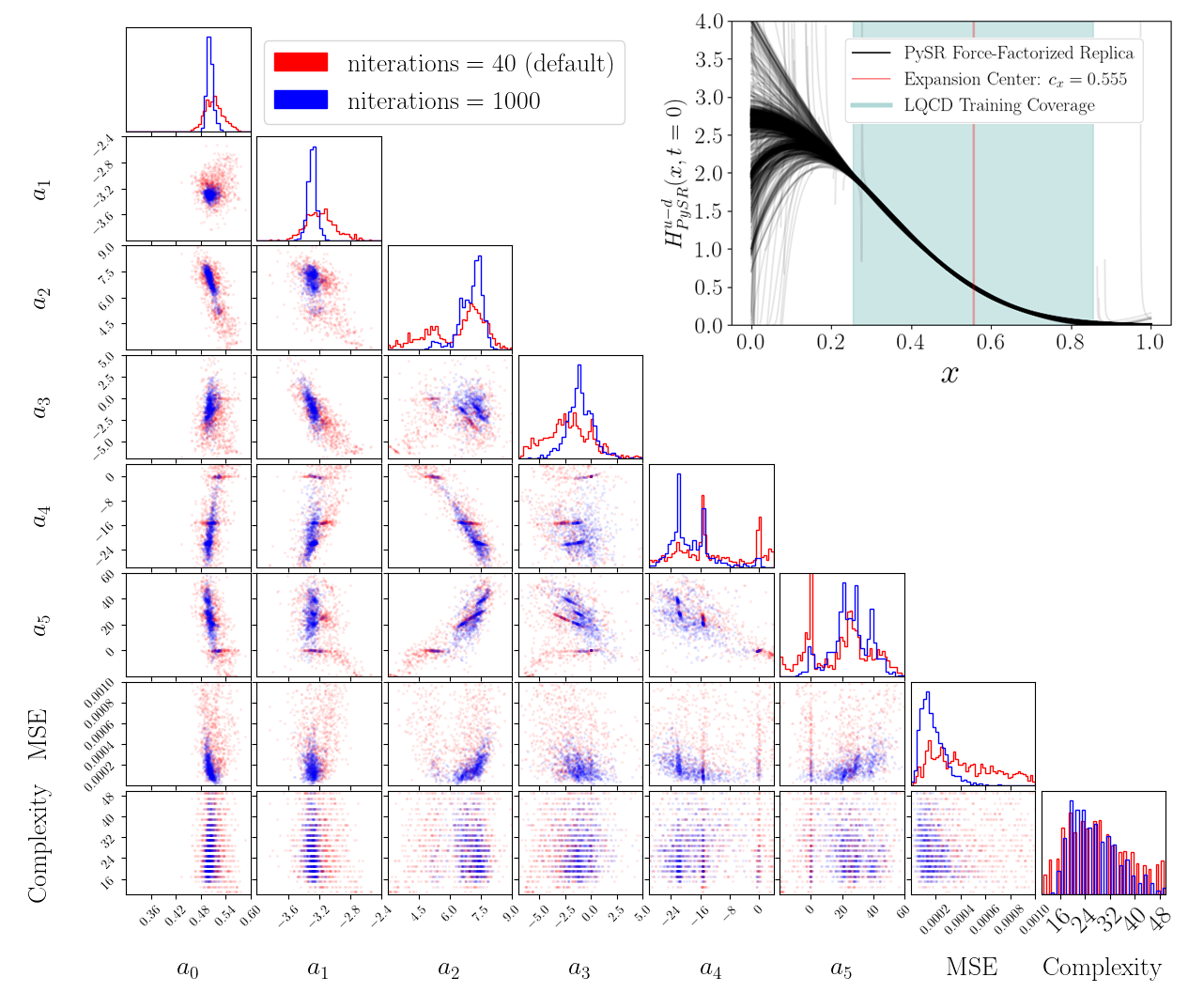}
\caption{Ensemble of 1000 factorized fits to the LQCD GPD data using PySR for $n_{iterations} = 40$ (default PySR setting, red) versus $n_{iterations} = 1000$  (blue).  The GPDs themselves for all 1000 replicas are plotted in the inset in the top-right corner (for $n_{iterations} = 1000$).  The array of smaller plots shows the distribution and correlations of Taylor coefficients $a_0 \dots a_5$, mean-square error (MSE), and complexity.  Clear clusters of preferred solutions are visible by eye, and increasing $n_{iterations}$ selects those preferred solutions more precisely. }
\label{fig: ForceFactorized_corner_spag_plot}
\end{figure}

\begin{figure}[H]
\centering
\includegraphics[width=0.4\linewidth]{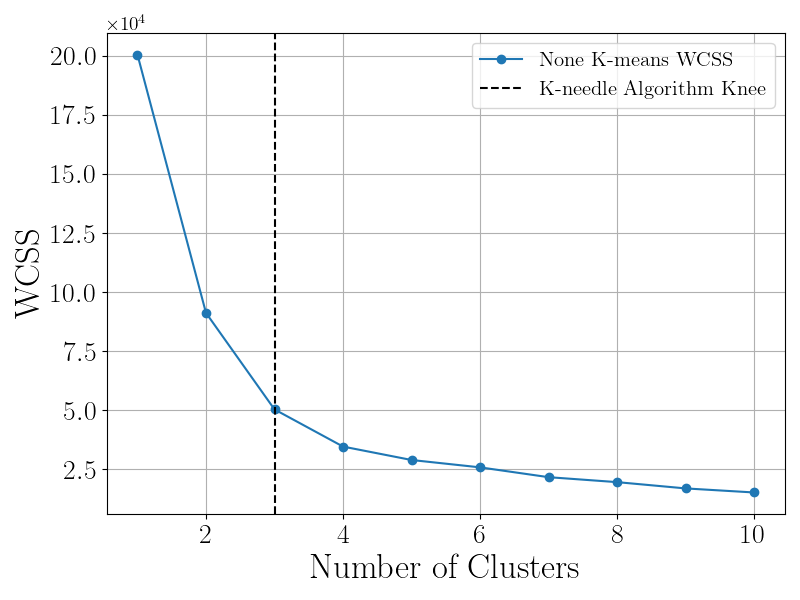}
\caption{K-means Within-Cluster Sum-of-Squares as a function of the number of clusters. The location of the elbow, selected by the K-needle algorithm, suggests that these FFreplicas are best classified into 3 clusters. 
}
\label{fig: FF_elbow_plot}
\end{figure}

For the Forced Factorization (FF) criterion, following the procedure described in Section \ref{subsubsec: BF_consistency}, we generated $10^3$ FF replicas PySR replicas and we performed an ECC (Section \ref{subsec:complexity}), on the distribution of replica Taylor coefficients in the PDF limit near $x=c_x = 0.555$ to $\ord{x-c_x}^5$. We see from Fig.~\ref{fig: ForceFactorized_corner_spag_plot} that most FF replicas describe the training region well, with some replicas containing poles in either the interpolation or extrapolation regions. In the low-x extrapolation region, we see the familiar clustering of replicas into bands, demonstrating that even in the absence of data at low x, PySR replicas largely extrapolate to only a few different behaviors. This is to say that, similar to the BF case, generating 1000 replicas does not result in 1000 different models, but only a few to within small deviations in the expansion coefficients.
Similar to the BF replicas, MSE alone was not sufficient for clustering these FF replicas. This is because PySR is converging towards replicas with comparable MSE, and therefore ECC or other choices for higher-dimensional clustering are necessary in order to adequately cluster the replicas. As in Section \ref{subsubsec: BF_consistency}, we apply Expansion Coefficient Clustering on the FF replicas to assign clusters to replicas that fall into the visible bands shown in Fig.~\ref{fig: ForceFactorized_corner_spag_plot}. We again use K-means algorithm to compute the WCSS per number of clusters, and fix the number of FF clusters using the K-needle algorithm (Fig.~\ref{fig: FF_elbow_plot}). We see from Fig.~\ref{fig: FF_elbow_plot} that the FF replicas are best classified into 3 clusters. 

\begin{figure}[H]
    \centering

    \begin{subfigure}[b]{0.45\textwidth}
        \centering
        \includegraphics[width=\textwidth]{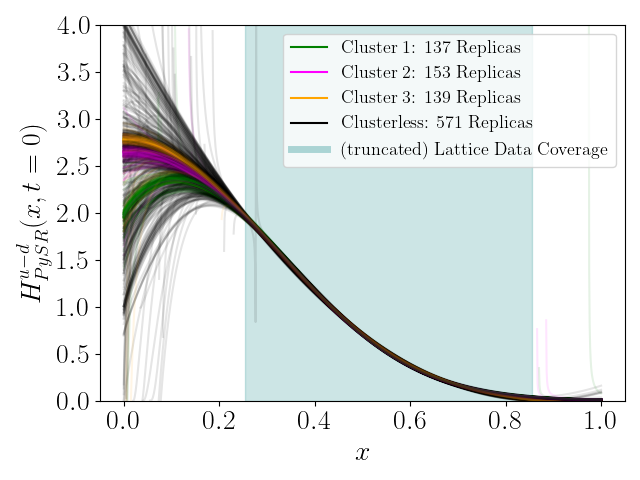}
        \caption{HDBSCAN clustering of 1k FFPySR replicas into 3 clusters based on location of K-needle knee.}
        \label{fig: FF_hwl_kneedle_spag}
    \end{subfigure}
    \hfill
    \begin{subfigure}[b]{0.45\textwidth}
        \centering
        \includegraphics[width=\textwidth]{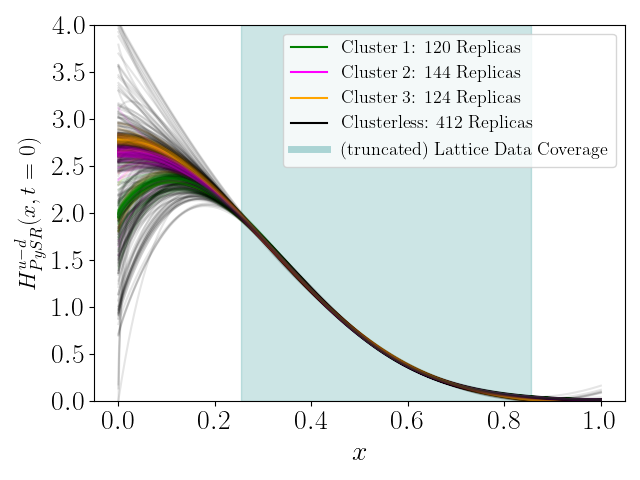}
        \caption{Remaining 800 FFPySR replicas (clustered with HDBSCAN and K-needle) after imposing Finite-Moment-Filter.}
        \label{fig: FF_hwl_good_kneedle_spag}
    \end{subfigure}

    \caption{Cluster Color-Coded spaghetti plot of FF before (left) and after (right) selecting onto replicas with finite $A_{10}(t=0)$ based on the K-needle algorithm  }
    \label{fig: FF_hwl_cluster_spag_plot}
\end{figure}

Due to the existence of FF replicas that extrapolate in-between the bands visible in Fig.~\ref{fig: ForceFactorized_corner_spag_plot}, we re-cluster the replicas using HDBSCAN, fixing $N_{\text{FF Clusters}}=3$ and include a clusterless category (Fig.~\ref{fig: FF_hwl_kneedle_spag}).
Though the behavior of FF replicas are similar to those of the BF replicas, the clustering of FF replicas is quite different. This is best understood after selecting onto FF replicas with finite $A_{10}(t=0)$ (Fig.~\ref{fig: FF_hwl_good_kneedle_spag}). Both sets of replicas (BF \& FF) were best described by 3 clusters using ECC, however, what was called Cluster 1 for the Best Fit replicas (Fig.~\ref{fig: BF_hwl_kneedle_spag}) is not present in the Force Factorized replicas. Manually fixing $N_{\text{FF Clusters}}= 4$ produces a cluster of FF replicas that does extrapolate similar to the BF Cluster 1 replicas, but this only results in 8 FF replicas after selecting onto replicas with finite $A_{10}(t=0)$ and so they are not considered here. That is, after filtering out models containing poles in $x$ at $t=0$, there is no visible FF cluster that corresponds to the low-x behavior of BF Cluster 1. Another interesting result of ECC applied to FF replicas is that though replicas within Cluster 2 and Cluster 3 extrapolate similarly at low-x, they are considered members of different clusters. This is no mistake as these two clusters correspond to replicas with different $a_5$. That is, these clusters describe genuinely different models that are degenerate at small-x.

\begin{figure}[H]
\centering
\includegraphics[width=0.6\linewidth]{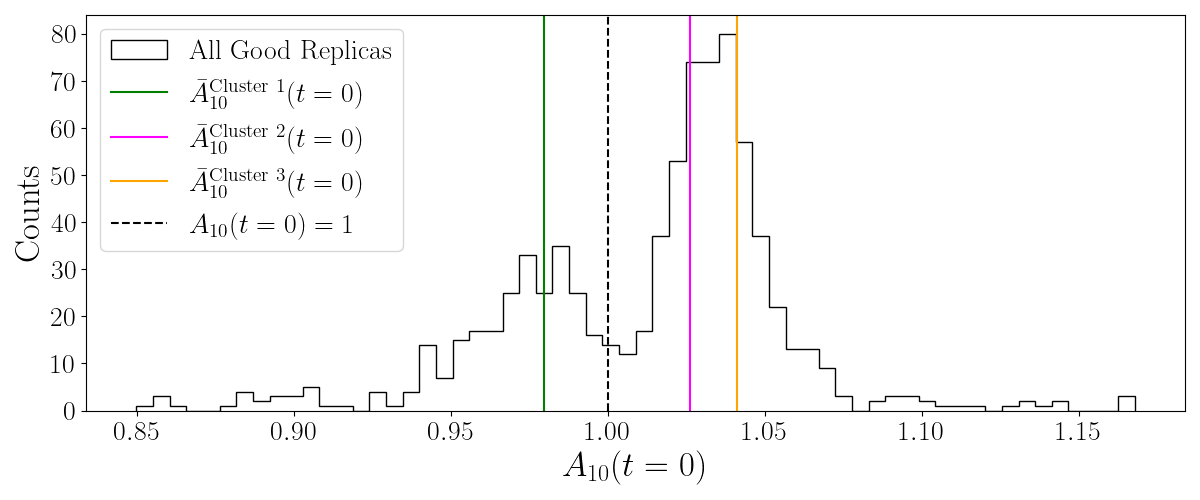}

\caption{Distribution of remaining $A_{10}(t=0)$ from FFPySR replicas. The average contribution to $A_{10}(t=0)$ from each cluster using K-needle is also shown by the color-coded vertical lines. The location of the peaks of $A_{10}(t=0)$ appear to be heavily correlated with cluster label, due to the location of the average value of $A_{10}(t=0)$ for a given cluster overlapping closely with the locations of the peaks of $A_{10}(t=0)$ for all 800/1000 remaining replicas. 
}
\label{fig: FF_1st_moment}
\end{figure}

\subsubsection{Semi-Reggeized}
\label{subsec: additional-approaches}
Semi-Reggeized forms, or RCA,  were analyzed by introducing the redundant variable $y\equiv 1-x$ so that $H_{u-d}=H_{u-d}(x,y,t)$, Eqs.\eqref{eq:reggeized_1} and \eqref{eq:reggeized_2}. 
For the semi-reggeized approach, we did not apply the series expansion and clustering. Instead we selected an exemplar (shown in Tab. \ref{fig:table}). We did however, run SR many times, and write an algorithm to check which results if any contained a Reggeized form without defining a custom loss that favors such forms. 
The exemplar selected is the first SR result obtained which satisfied the Reggeized format, 
\begin{equation}
\label{eq: regge-exemplar}
    H_{u-d}^{Regge}(x,{\bf t}) = 2.74 \,\, e^{-\tcal} \, 
    x^{0.087} 
    (1-x)^{[1.57 - 0.67 \, \text{ln}(1-x)]}
\end{equation}
where $\tcal \equiv -t/\Lambda^2$, and $\Lambda^2 = 1$ GeV$^2$. The selected Reggeized exemplar has MSE = $1.9\times 10^{-4}$ relative to the testing dataset. Though the MSE is notably larger than that of the BF or FF LQCD exemplars, the MSE for the Reggeized exemplar is the same order of magnitude as the average MSE of FF LQCD replicas (Tab. \ref{tab:mse_stats}), a testament to the convergence of PySR with well chosen hyperparameters. We from Eq. \eqref{eq: regge-exemplar} that along with the characteristic Regge behavior $x^\alpha(1-x)^\beta$ for $H^{Regge}_{u-d}(x,\tcal)$, there is residual x-dependence in the exponential that vanishes as $x\rightarrow 0$. Though $\ln^2(1-x)$ diverges at $x= 1$, $\exp(
-\ln^2(1-x))$ vanishes as $x\rightarrow 1$, recovering the characteristic Regge behavior at both endpoints. In future work, we may devise a 'Force-Reggeized' (FR) custom loss function and carry out a procedure such as we did in ECC as a way to test the consistency, convergence, and clustering of Reggeized SR replicas
The KL-divergence of an FR-custom-loss KL(MSE$_{BF}||$MSE$_{FR}$), would allow one to make more quantitative statements regarding the truth of the Reggeized hypothesis for a given dataset. It is of course reassuring that the \textit{Redundant Column Approach} in conjunction with post-facto selection (based on functional form, not MSE) resulted in a Reggeized model that performs comparably well to the average FF replica.

\subsubsection{Uncertainty Treatment}
\label{subsec:uncertain}

The uncertainty in SR determinations can appear in two distinct instances in this work: 1) point-by-point data uncertainty and 2) global uncertainty introduced by the nature of SR dependent upon genetic  annealing evolution (Section \ref{sec:3}). Both sources of uncertainty are reflected in the final choice of each predicted curve shown in our spaghetti plots (Figs. \ref{fig: BF_hwl_kneedle_spag} \ref{fig: FF_hwl_kneedle_spag}).

To handle the point-by-point uncertainty,
in this work we adopt MSE Eq.\eqref{eqn:MSE} for the "mean of squared residuals" and WMSE Eq.(\ref{eqn:WMSE}) as the "weighted mean squared residuals" where residuals are defined as difference between the  the variable's observed  and predicted values, for each given input data point.  However, it should be remarked that the LQCD errors are highly correlated but unaccounted for. LQCD Correlations are visibly apparent from their trend in both the kinematic variables, $x$ and $t$ and are due to the dominance of systematic errors over the statistical ones (for an explanation of LQCD systematics for this calculation see \cite{Lin:2020rxa}. Therefore, in our analysis, while adopting the ``MSE" and ``WMSE" terminology from the SR literature, it is understood that these quantities do not carry statistical significance. They are, however, used as ``fitness metrics", to gauge the goodness of the various symbolic expressions approaching the LQCD curves. Values of fitness metrics for exemplar curves are presented in Tab. \ref{fig:table}. 

To handle global uncertainty introduced by SR, we assess the aggregate of many SR attempts using the same data, and find that SR converges in the training region and introduces clusters of solutions in the extrapolation region. 
These clusters of solutions are introduced by SR while attempting to perform a global optimization of the  fitness metric, which is not changing. The same global optimum yields multiple clusters of solutions using the same SR algorithm many times.

In this work, we see that clustering of SR results show tension between two effects. 
The space of possible curves widens, and the number of clusters can go up, as the uncertainty on the data increases, and SR has access to more composite functions, and physical constraints are ignored.
Conversely, as physical constraints are applied, data-uncertainty shrinks, and the number of composite functions is reduced than the space of possible result curves from SR shrinks, and the number of viable clusters decreases.
Specific examples of physical constraints include but are not limited too polynomiality, positivity, matching onto form factors/PDFs, normalization.  
We reserve treating the correlated uncertainties, more statistical MCMC SR approaches, and additional physical constraints for future work.

\begin{figure}[H]
\centering
\includegraphics[width=0.50\linewidth]{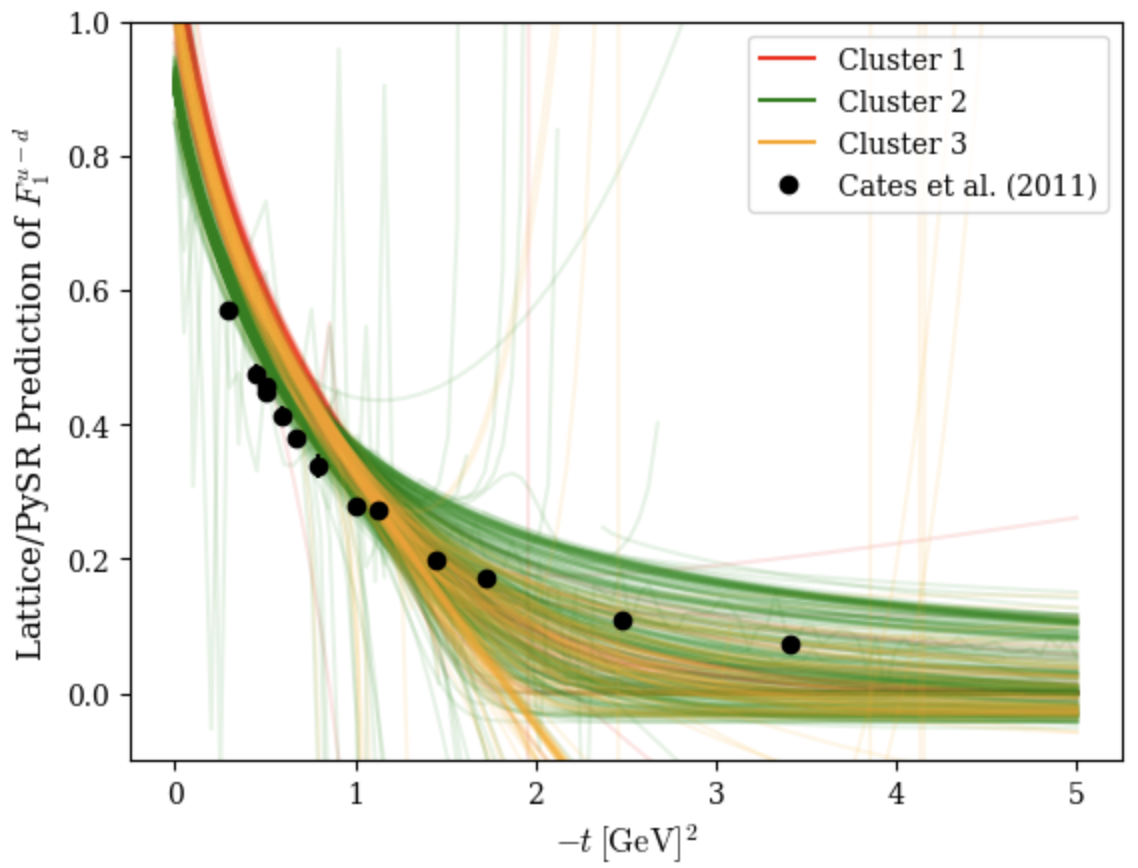}
\caption{BF SR results for the flavor separated electromagnetic form factor, $F_1^q$, Eq.\eqref{eq:mom1}, with $q=u-d$, calculated from Fig. \ref{fig: BF_hwl_kneedle_spag}. Experimental data from Ref.\cite{Cates:2011pz}. Within the region of the training data, $0.1 \mathrm{\: GeV^2} \leq \mid t \mid \leq 1 \: \mathrm{GeV^{2}}$, the symbolic expressions track the behavior of the form factor.}
\label{fig: FF_check}
\end{figure}

Because a single SR run, produces a global minimum curve and parameters we do not have a statistically sound way of producing a spaghetti plot. Running SR many times, produces many best fit curves, and we can assess how SR introduces an intrinsic spread of its own results, in other words, we can perform, in this case a statistical analysis, and define an error band for the results. This is illustrated  in Fig.~\ref{fig: FF_check} where we plot the integrated over $x$ GPD, $H_{u-d}$, namely the Dirac form factor, $F_1^{u-d}$, Eq.\eqref{eq:mom1}, vs. $-t \equiv \mid t \mid$. 
The figure clearly shows the spread in the spaghetti plot that can eventually be cast into a band around the most likely result, however the band does not correspond to a clearly defined probability. 
Note that the integration over $x$ extends beyond the region of the training data. Therefore the agreement with data demonstrates that our model extrapolates well  in both $x$ and $t$.


\subsection{PySR Selected Model Results}
\label{subsec:PySR_results}
As shown in Section \ref{subsec:results}, we were able to map the various extrapolating behaviors of PySR replicas to clusters using ECC. 
In Fig.~\ref{fig:table}, we present a typical result for each cluster. The selected LQCD $H_{u-d}(x,t)$ PySR forms are organized as $H^{i}_{BF}(x,{\bf t})$, $H^{i}_{FF}(x,{\bf t})$, $i=1,K$ taking a Kneedle number of clusters $K= 3$ (Sections \ref{subsubsec: BF_consistency}, \ref{subsubsec: FF_consistency}), and $H_{Regge}$. 
The specific forms of $H^{i}_{BF}(x,{\bf t})$ ($BF$=Best-Fit), or $H^{i}_{BF}(x,{\bf t})$, ($FF$=Force-Factorized), are chosen as the ones that give the lowest MSE in each category, where ${\bf t}$, is a dimensionless quantity ${\bf t} \equiv \mid t \mid/\Lambda^2$, with parameter $\Lambda^2 = 1 $ GeV$^2$. 
Furthermore, by selecting the replica with the lowest MSE for that cluster, that also has finite $A_{10}(t=0)$. 
The second column of the Table illustrates their 3D graphic form, while the math expressions are shown in column 3.


The last two columns illustrate the fitness metric, explained in Section \ref{subsec:uncertain}, and the complexity, respectively.

\begin{table}[H]
\centering
\includegraphics[width=0.9\linewidth]{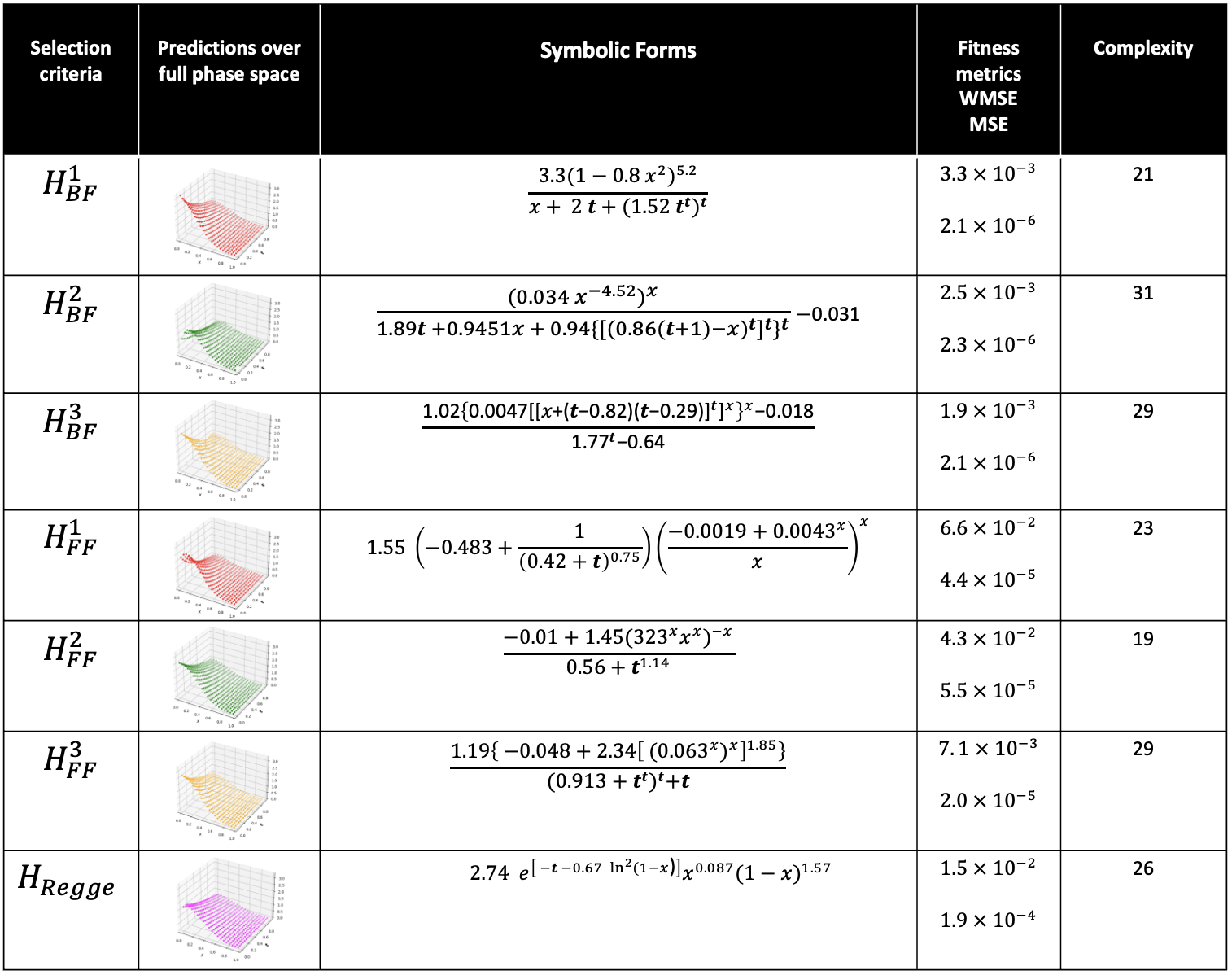}
\caption{SR functional forms for all three models: Best-Fit (BF), Forced-Factorized Fit (FF), and Semi-Reggeized (Regge). For BF and FF, the functional forms corresponding to the three different clusters determined in this analysis are displayed and numbered, with $K=1,3$. The fitness metric and complexity values associated with each determination are discussed in the text.  The boldface $\textit{\textbf{t}} = {-t}/{\Lambda^2}$, with $\Lambda^2 = 1$ GeV$^2$,  is used to ensure that all numerical coefficients in the symbolic expressions are dimensionless.
}
\label{fig:table}
\end{table}

The table also illustrates the PySR Exemplar Complexities obtained before simplifying w/ sympy.

$H_{Regge}$ was determined with the RCA method described in Section \ref{subsec: additional-approaches}. We see that along with the characteristic Regge behavior $x^\alpha(1-x)^\beta$, there is residual $x$-dependence in the exponential. Eulers constant $e$ showing up as the base of the exponential is an artifact of the fit being performed in log space and then exponentiated, as opposed to PySR serendipitously discovering $e$. 
Though the MSE of this Reggeized exemplar is notably larger than that of the BF or FF LQCD exemplar, it is the same order of magnitude as the average MSE of FF LQCD replicas (Tab. \ref{tab:mse_stats}). 
In future work, we will devise a ``Force-Reggeized" (FR) custom loss function and carry out the procedure of ECC in order to test the consistency and convergence of Reggeized SR replicas, analogous to the treatment of the Best Fit and FFreplicas. 
%

\subsection{How reliably can we test factorization?}
\label{subsec: test_factorization}
As noted in Section \ref{sec:2}, one can glean information on the nucleon quark and gluon spatial substructure by studying how configurations of a given $x$, depend on $t$, and consequently, by Fourier transformation, how they are distributed in the spatial variable ${\bf b}$ around the proton's relativistic center of momentum. An obvious case is when the $x$ and $t$ behaviors are independent of each other, describing the GPD in a factorized form that
leads to a constant RMS radius for all values of $x$ (see Eq.\eqref{eq:radius}).

To study this problem, phenomenological parametrizations in the variables $x$ and $t$, have been constructed using known theoretical information on the symmetry constraints and specific functional behavior. 
For instance, due to the constraint, $H_q(x,0,0) = q(x)$, Eq.\eqref{eq:pdf}, the GPD $x$ behavior in regions, $x \rightarrow 0$ and $x \rightarrow 1$, can be assumed to be similar to  the $x^{-\alpha}$ or Regge behavior, and $(1-x)^\beta$, power-like behavior obtained from quark counting rules, respectively (we refer the reader to Refs.\cite{Landshoff:1970ff} and \cite{Sivers:1982wk} for reviews). At intermediate $x$, interpolated forms can then be derived, preserving the smooth shape of the distributions. 
The $t$ dependence for these parametrizations is also modeled along similar lines, in this case, using besides the predicted Regge behavior in $t$, the constraint from  Eq.\eqref{eq:mom1}. 

For our study of LQCD results, we use three phenomenological GPD parameterizations for the quantity, $H_{u-d} = H_u-H_d$, as benchmarks for the performance of   
our ansatz in PySR. The chosen parametrizations are labeled as: VGG \cite{Vanderhaeghen:1999xj}, GGL \cite{Kriesten:2021sqc}, and GK \cite{Goloskokov:2005sd}, respectively.
GGL and GK  are constructed so that they explicitly break factorization in $x$ and $t$, whereas VGG exhibits approximate factorization. 
The behavior of the distributions concerning factorization can be shown by plotting the ratio $R(x,|t|)$, 
\begin{equation}
    R(x,|t|) = \frac{H_{u-d}(x,0)}{H_{u-d}(x,|t|)}  , 
    \label{eq:factorization}
\end{equation}
where, if the source/model factorizes, $R$ is constant at fixed $|t|$. 


\begin{figure}[H]
    \centering  
    \textbf{Fixed $|t|$ ratios $R(|t|)$}\par\medskip
    
\begin{subfigure}[b]{0.45\textwidth}
    \centering
    \includegraphics[width=0.8\textwidth]{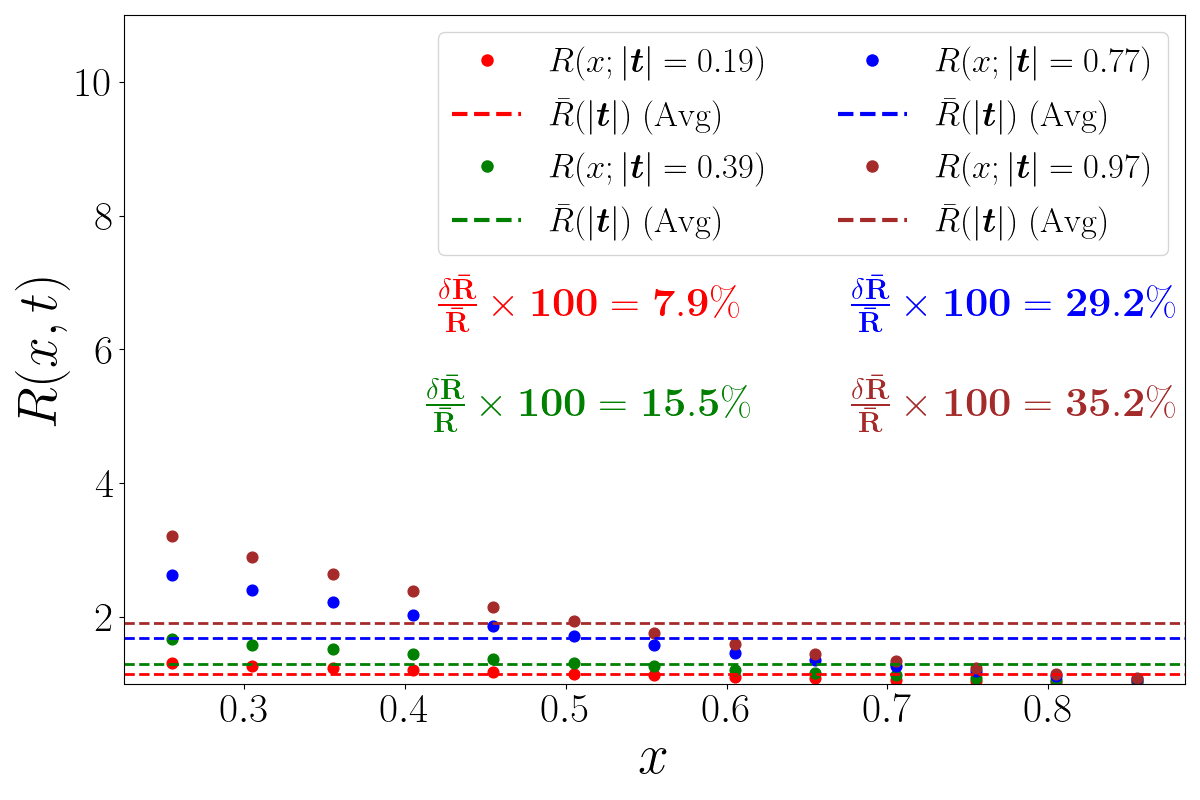}
    \caption{GGL}
    \label{fig: ggl_ratios}
\end{subfigure}
\hspace{-0.5cm}
\begin{subfigure}[b]{0.45\textwidth}
    \centering
    \includegraphics[width=0.8\textwidth]{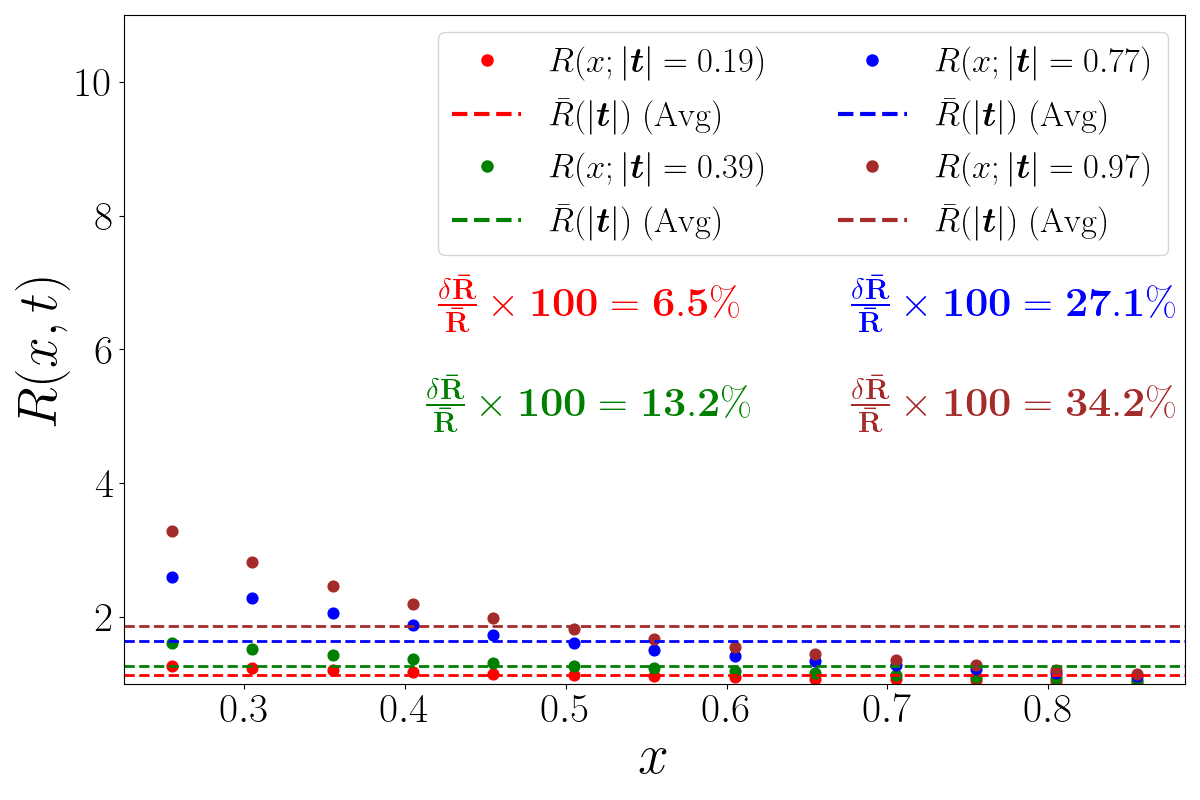}
    \caption{GK}
    \label{fig: gk_ratios}
\end{subfigure}

\vspace{0.5cm}

\begin{subfigure}[b]{0.45\textwidth}
    \centering
    \includegraphics[width=0.8\textwidth]{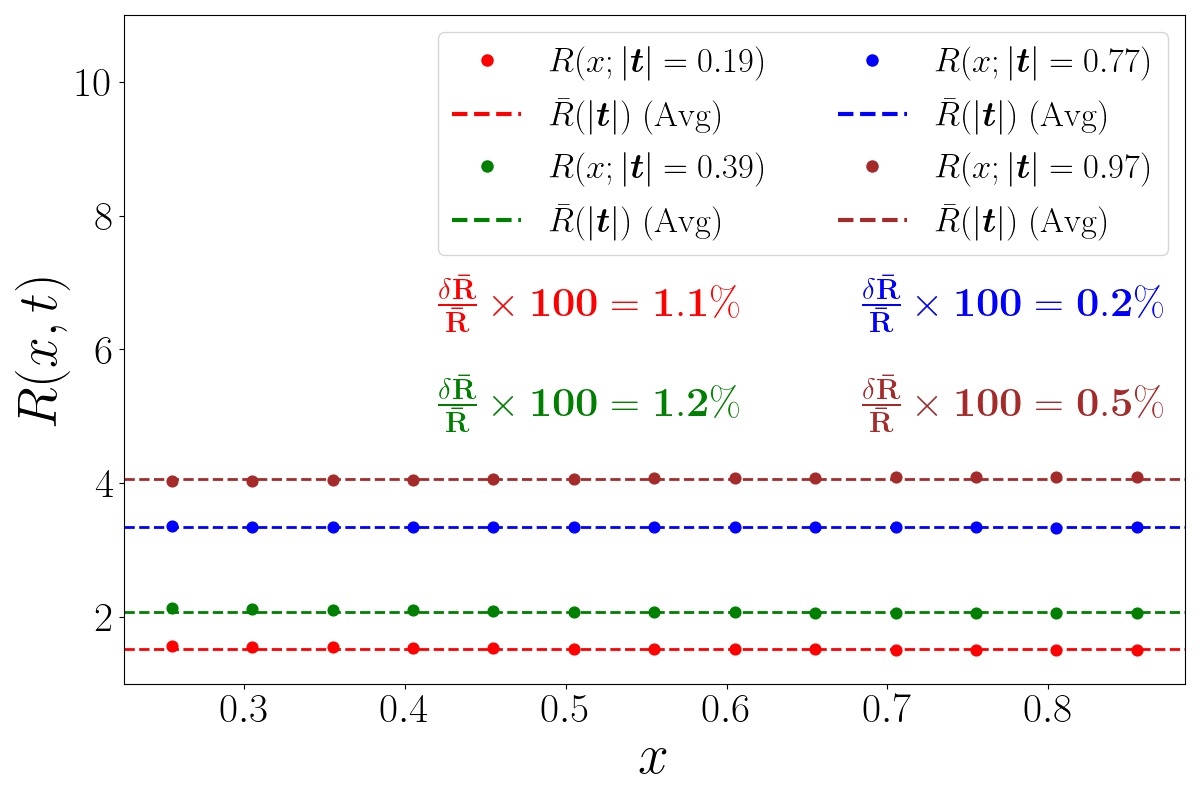}
    \caption{VGG}
    \label{fig: VGG_ratios}
\end{subfigure}
\hspace{-0.5cm}
\begin{subfigure}[b]{0.45\textwidth}
    \centering
    \includegraphics[width=0.8\textwidth]{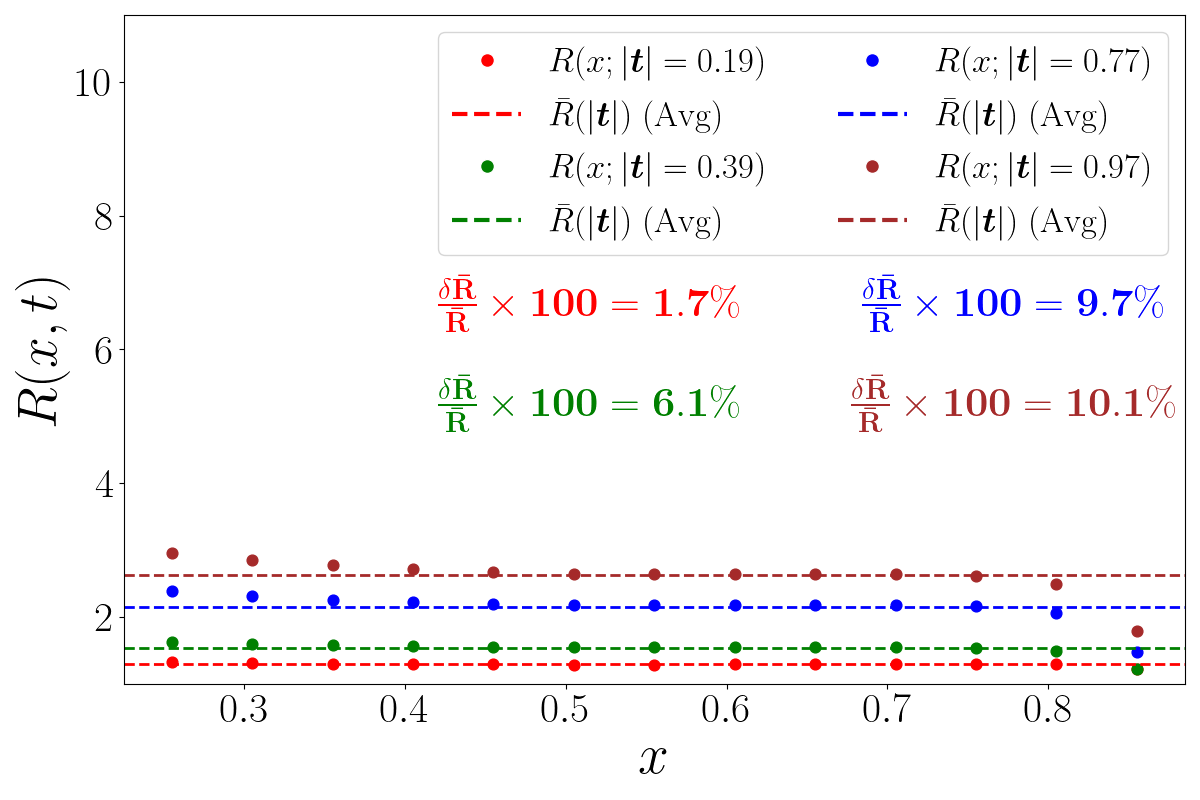}
    \caption{LQCD}
    \label{fig: Lin21_ratios}
\end{subfigure}

    \caption{Ratios $R(x,|t|)$ of GPD parameterizations VGG, GGL, GK as well as LQCD results \cite{Lin:2021brq} plotted vs. $x$ for different values of $|t|: 0.19, 0.39, 0.77, 0.97$ GeV$^2$.  The $x$ dependence of the GGL and GK ratios demonstrates the factorization breaking in these parameterizations. VGG, on the other hand, explicitly factorizes for any quark flavor, and approximately factorizes in the flavor isovector combination. Similarly, we see approximate factorization in the LQCD in this region, in particular in the interval $0.45\leq x\leq 0.75$. No symbolic regression (SR) is used to produce the content of this figure.}
    \label{fig: source_ratios}
\end{figure}
The ratio $R$ is plotted vs. $x$ for various values of $|t|$, in Fig.~\ref{fig: source_ratios}, for the phenomenological parametrizations as well as the LQCD results.
The approximate factorized behavior of the VGG model is clearly visible as compared to the behavior of GGL and GK which deviate from constant by over $30\%$ at $|t|\approx0.97$ GeV$^2$, suggesting that in an SR analysis they would be poorly described by a FF model. 
For the LQCD source data -- for which no analytic expression is known -- we find that approximately factorizes to within $\approx 10\%$ in the training region. However,  the percent deviation of LQCD from constant is larger than that of VGG. 

\begin{figure}[H]
\centering
\includegraphics[width=0.35\linewidth]{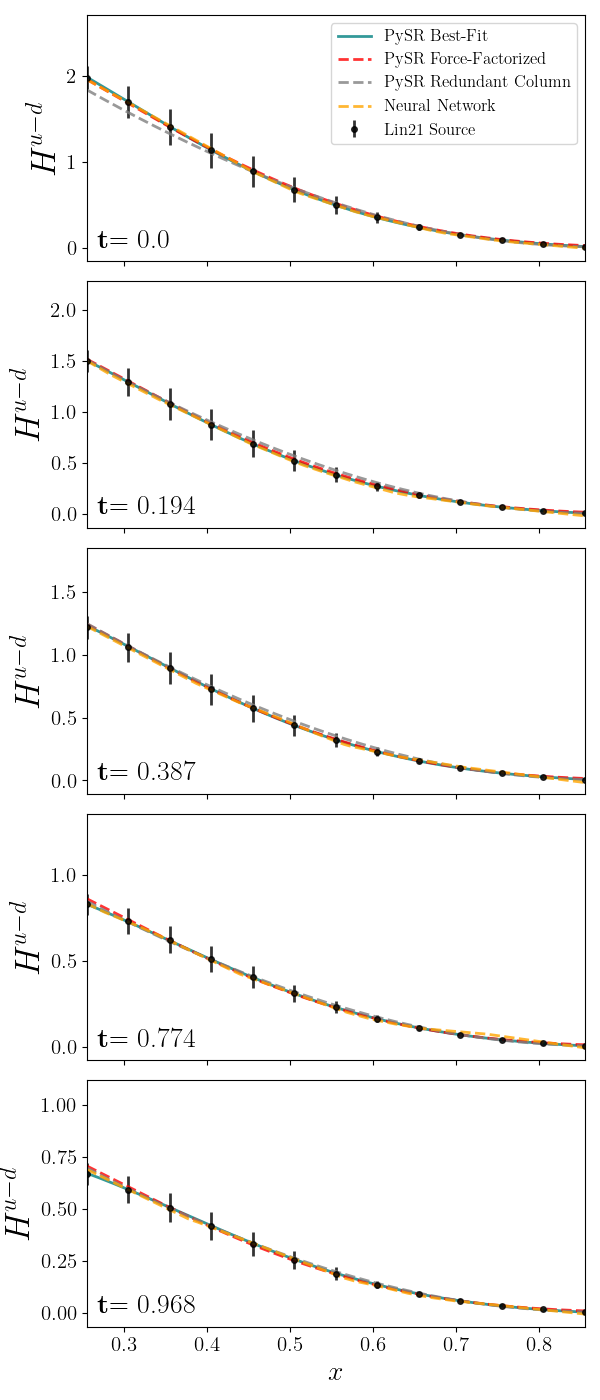}
\caption{Array of fits to the lattice source data \cite{Lin:2020rxa} for $H^{u - d}(x,\tcal)$ projected onto fixed $\tcal\equiv -t/ \Lambda^2~$, with $\Lambda^2=1$ GeV$^2$. The PySR Best-Fit, Force-Factorized, and Redundant Column models shown correspond to $H^3_{BF}$, $H^2_{FF}$, and $H_{Regge}$ in Tab. \ref{fig:table}, respectively.}
\label{fig: HWL_5panel}
\end{figure}

\begin{figure}[H]
\centering
\includegraphics[width=0.32\linewidth]{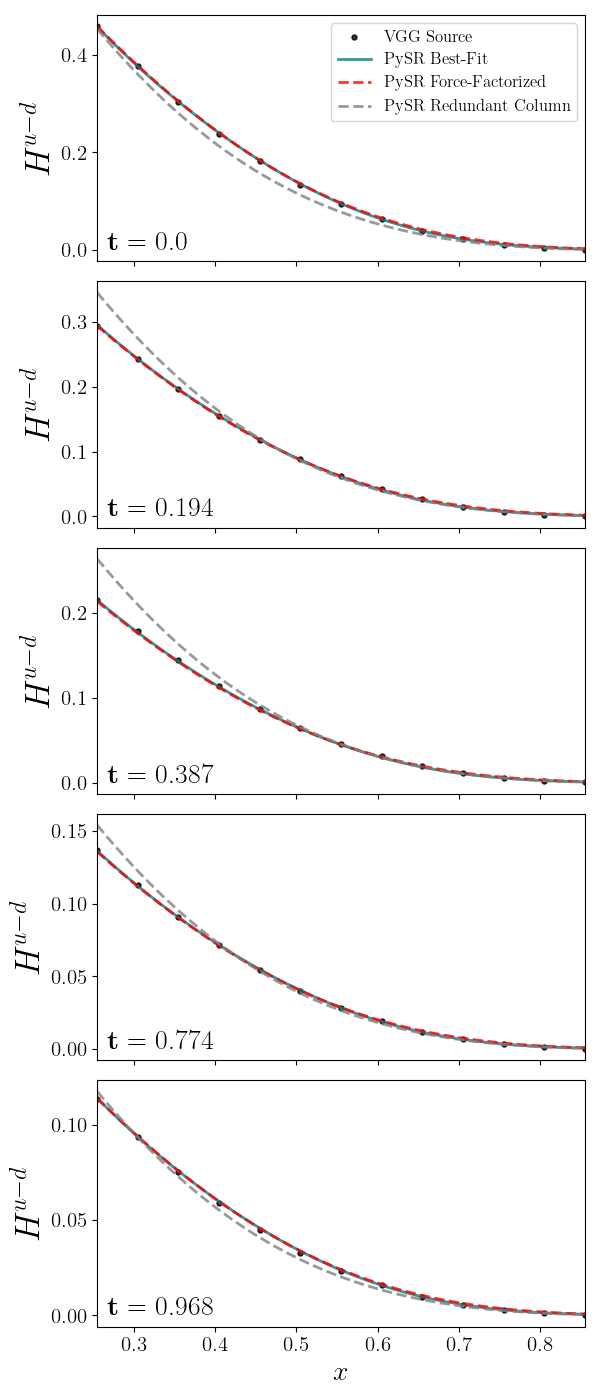}
\includegraphics[width=0.32\linewidth]{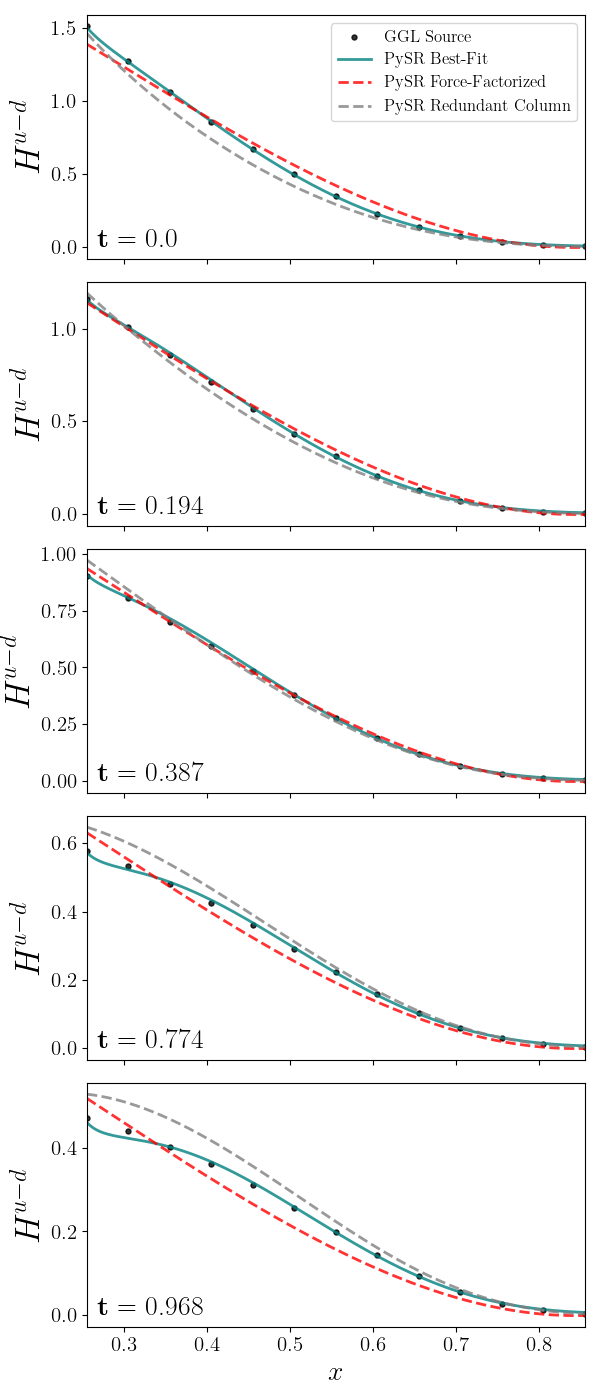}
\includegraphics[width=0.32\linewidth]{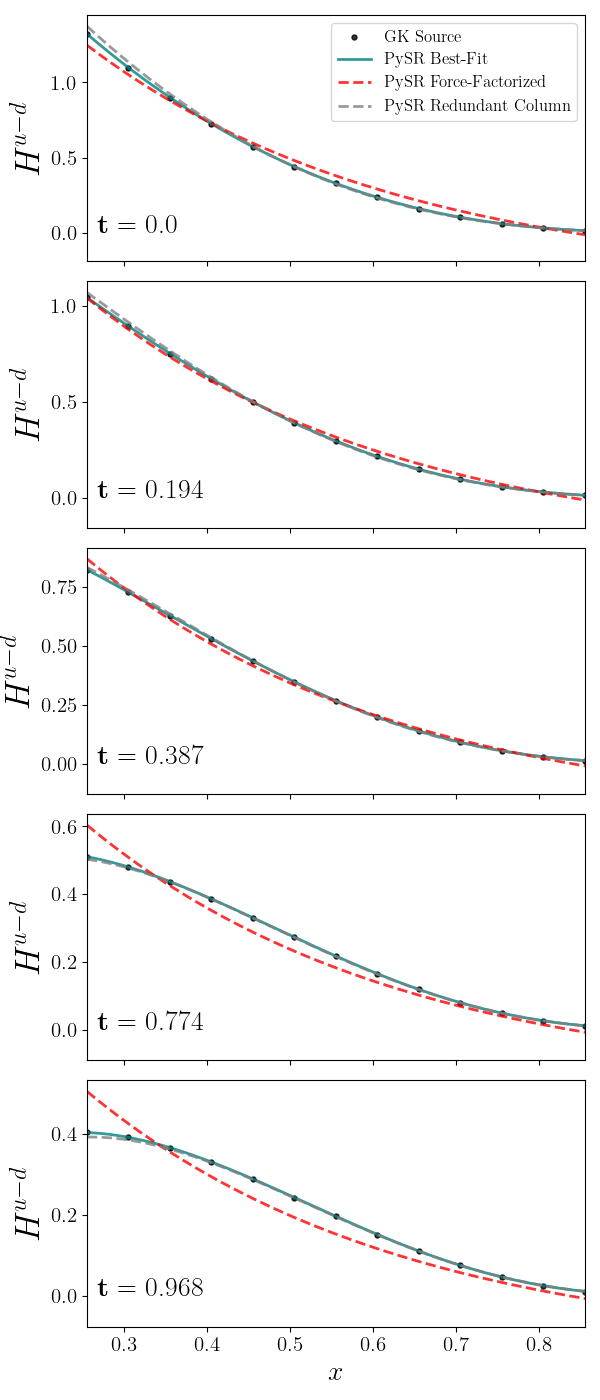}
\caption{Array of fits to the VGG \cite{Vanderhaeghen:1999xj} (left) GGL \cite{Kriesten:2021sqc} (center) and GK \cite{Goloskokov:2005sd} (right) source data for $H^{u - d}(x,\tcal)$ projected onto fixed $\tcal\equiv -t/ \Lambda^2~$, with $\Lambda^2=1$ GeV$^2$. Best-Fit and Force-Factorized PySR replicas with the lowest WMSE were selected as exemplars without applying the clustering procedure from Sec. \ref{subsubsec: BF_consistency} and are omitted from Tab. \ref{fig:table}.
 }
\label{fig: phenom_5panel}
\end{figure}

In Fig.~\ref{fig: HWL_5panel} and \ref{fig: phenom_5panel} we show the exemplars from each cluster presented in Section \ref{subsec:results} plotted against LQCD results and the phenomenological models, respectively. The curves in the figures were chosen based on which replica had the lowest MSE per cluster, representing the BF, FF, and Semi-Reggeized models. In each panel we plot the various functional forms vs. $x$, at the $t$ values calculated in LQCD. 
In the LQCD case (Fig.\ref{fig: HWL_5panel}), we also compare our SR results to that of a feed-forward neural network, implemented in the PyTorch library \cite{paszke2019pytorchimperativestylehighperformance}. 
Considering our dataset is relatively small and low-dimensional, we only require the use of one hidden layer. 
The popularity of the use of ML for scientific discovery owes much of its rise to the effectiveness of neural networks, which have constituted a large part of the hadronic physics community's embrace of ML techniques, e.g. \cite{Kumericki:2011rz, Nocera_2014, NNPDF:2014otw, Moutarde_2019, PhysRevD.104.016001}. Although a powerful tool for regression and classification, NNs are plagued with issues of interpretability due to the ``black box" nature of their hidden layers. 
On the other hand, the major advantage of symbolic regression is that the learned model is interpretable; as we demonstrate throughout this work, SR oftentimes gives human-readable expressions from which one can immediately extract physics content. 
%
While symbolic regression and neural networks both provide excellent fits to the data, as is evident from Fig. \ref{fig: HWL_5panel}, we can readily interpret the output of the former and not the latter. 



%

\vspace{0.2cm}
To test PySRs capacity to falsify various physics assumptions in the SR calculated functional forms behavior, such as factorization, we first define the \textit{Kullback-Leibler (KL) Divergence} \cite{KL-divergence} for two probability distributions $P$ and $Q$ evaluated at points $i\in X$:
\begin{equation}
    \text{KL}(P||Q) = \sum_{i\in X}P_i\log\frac{P_i}{Q_i}
    \label{eq:KL}
\end{equation}
The KL-divergence is a statistic for quantifying the amount of information lost when replacing distribution $P$ with $Q$. Bound between $0\leq \text{KL}(P||Q)\leq \infty$, the lower limit is saturated when the two distributions are identical ($P_i=Q_i~\forall ~i\in X$ ), and the upper limit is saturated when the same measurement from two distributions are incompatible (Eg. $P_1 \neq 0$ and $Q_1=0$).

We use Eq.\eqref{eq:KL} to compare the MSE distributions from our PySR analysis, applied to all four sources: LQCD and the three models (VGG, GGL and GK), for both BF and FF custom loss functions. 
We generate $10^2$ PySR replicas for each source and we extract the distribution of MSEs relative to the testing data 
selected using the equivalent procedure shown in Fig.~\ref{fig: PySR_HWL_train_test} 
for each set of replicas. 
For each source we then compute the KL-divergence of the BF MSEs from the FF MSEs,
\begin{equation}
    KL(\text{MSE}_{\text{BF}}||\text{MSE}_{\text{FF}})
    \label{eq:KL_MSE}
\end{equation}
through the Python package scipy.stats.entropy\cite{KL-entropy}.
Our baseline for defining two compatible distributions is provided by the VGG source that factorizes to $\approx 1\%$ (Fig.\ref{fig: source_ratios}), 

\[KL(MSE^{\text{VGG}}_{\text{BF}}||\text{MSE}^{\text{VGG}}_{\text{FF}}) \]
%
In Table \ref{tab:mse_stats} we compare 
the separation between VGG KL-divergence and the ones for LQCD, GGL, and GK source MSEs.
\begin{table}[H]
\centering
\renewcommand{\arraystretch}{1.4}
\begin{tabular}{|p{3.2cm}||p{2.5cm}|p{2.5cm}|p{2.5cm}||p{2.5cm}|}
\hline
\multicolumn{5}{|c|}{PySR Fit Statistics} \\
\hline \hline
 & LQCD & VGG & GGL & GK \\
\hline
MSE (BF) & $6.86 \times 10^{-5}$ & $1.55 \times 10^{-6}$ & $5.18 \times 10^{-5}$ & $8.35 \times 10^{-6}$ \\
MSE (FF) & $1.77 \times 10^{-4}$ & $5.10 \times 10^{-6}$ & $1.87 \times 10^{-3}$ & $1.70 \times 10^{-3}$ \\
MSE Ratio: FF/BF & $2.25 \times 10^{0}$ & $3.29 \times 10^{0}$ & $3.61 \times 10^{1}$ & $2.03 \times 10^{2}$ \\
\hline \hline
KL  & $3.0 \times 10^{-2}$ & $2.68 \times 10^{-2}$ & $1.86 \times 10^{-1}$ & $6.17 \times 10^{-1}$ \\
\hline
\end{tabular}
\caption{Average MSE of Best-Fit (BF) and Force-Factorized (FF) PySR replicas calculated on the testing dataset. The KL divergence being computed is 
 $KL ($Best-Fit$||$Force-Factorized$)$ (Eq.\eqref{eq:KL_MSE}). 
}
\label{tab:mse_stats}
\end{table}

%
We see from Tab. \ref{tab:mse_stats} that 
for all datasets for which the MSE was calculated, the corresponding KL-divergences are consistent with VGG and LQCD approximately factorizing, while both GGL and GK are described poorly with a FF model. 
%
%

\vspace{0.2cm}
\noindent {\it VGG and LQCD}

\noindent Interestingly, the LQCD results outperformed the ``truth" case of VGG, with lower values for both the MSE ratio and KL-divergence. Though the LQCD source data itself breaks factorization more than VGG (Fig.~\ref{fig: source_ratios}), LQCD results for the MSE ratio and KL-divergence still show that LQCD is compatible with factorization on the region. There are many factors that could cause LQCD to outperform the truth case, such as the source data for VGG and LQCD not being identically easy for PySR to learn. This is suggested by the near order-of-magnitude larger value of MSE$^{\text{LQCD}}_{\text{BF}}$ compared to that of VGG. 
That increase in BF MSE closes the gap more between MSE$^{\text{LQCD}}_{\text{BF}}$ and MSE$^{\text{LQCD}}_{\text{FF}}$, artificially making it appear as though LQCD factorizes better than VGG, when in reality BF LQCD results simply converged less than BF VGG. 

\vspace{0.2cm}
\noindent{\it GGL and GK}

\noindent Both GGL and GK KL-divergences suggest incompatibility with factorization, however the difference between the two KL-divergences naively suggests that $GGL$ is more compatible with factorization than $GK$. This seemingly contradicts the percent deviations provided in Fig.~\ref{fig: source_ratios}, which show that that GGL source data breaks factorization slightly more than GK. However, as was the case when comparing VGG and LQCD, MSE$^{\text{GGL}}_{\text{BF}}$ is notably larger than that of $VGG$, resulting in the apparent contradiction. For this reason, we are unable to establish hierarchy of source factorization breaking based on these results, but we are able to falsify the assumption that a given source factorizes in general.

\begin{figure}[H]
\centering
\includegraphics[width=0.40\linewidth]{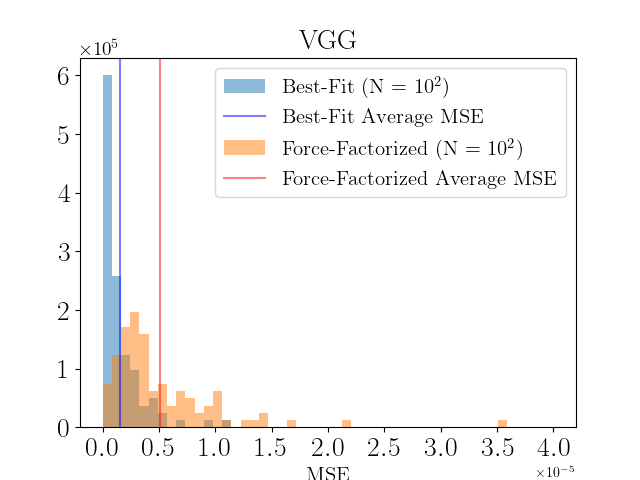}
\includegraphics[width=0.40\linewidth]{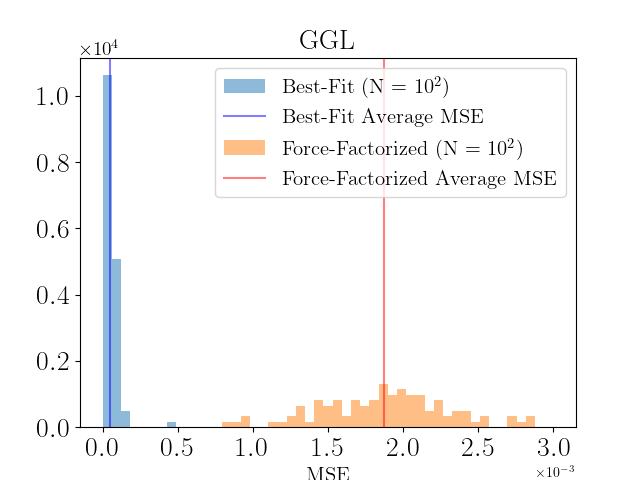}
\includegraphics[width=0.40\linewidth]{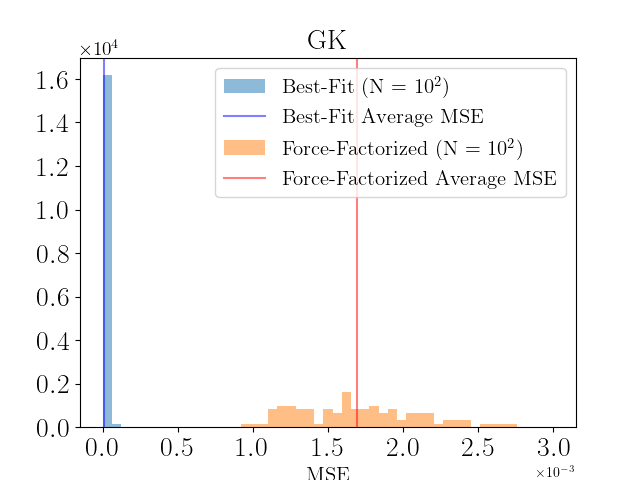}
\includegraphics[width=0.40\linewidth]{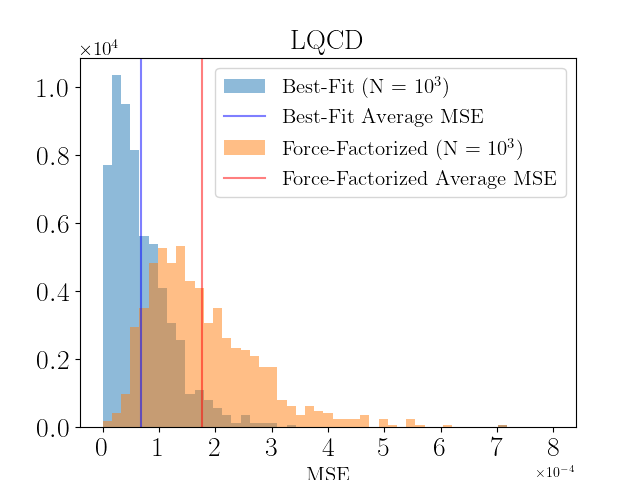}
\caption{Histogram of the MSEs relative to the testing dataset for 100 BF and FF replicas trained on Phenomenological sources (VGG, GGL, and GK) as well as lattice source data \cite{Lin:2020rxa}. All MSE distributions are normalized to one. The overlap in MSEs for the VGG case is consistent with VGG being known to approximately factorize, while the clear separation of MSE for GGL and GK between BF and FF replicas is consistent with these sources being incompatible with the factorization ansatz. 
Note that the areas shown in these histograms are not to scale, because of the log-scale MSE-axis.
}
\label{fig: MSE_separation}
\end{figure}

\vspace{0.2cm}
\noindent  
In Fig.˜\ref{fig: MSE_separation}, we present histograms of the MSEs relative to the testing dataset for the BF and FF replicas trained on LQCD, VGG, GGL, and GK,
to illustrate how the VGG and LQCD sources are compatible with factorization. This is shown in the graph by the overlap of MSEs for both BF and FF replicas. 
On the other side, we also see a clear lack of overlap between BF and FF MSEs for GGL and GK sources, implying that they are not compatible with the factorization hypothesis.

In future work, one might consider generating two sets of BF replicas for each source instead of one, and computing the corresponding KL-divergence KL(\text{MSE}$^1_{\text{BF}}||\text{MSE}^2_{\text{BF}})$ as a baseline for compatible distributions. This may help eliminate bias due to comparing KL-divergences across sources, rather than the comparison being across different assumptions for the same source.

\subsubsection{LQCD Moments}
\label{sec:moments}
\begin{figure}[H]
    \centering
\includegraphics[width=0.475\textwidth]{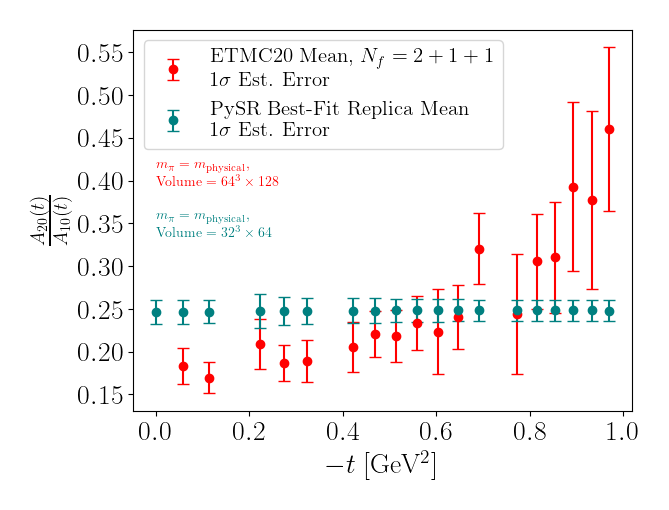}
    \hfill
\includegraphics[width=0.475\textwidth]{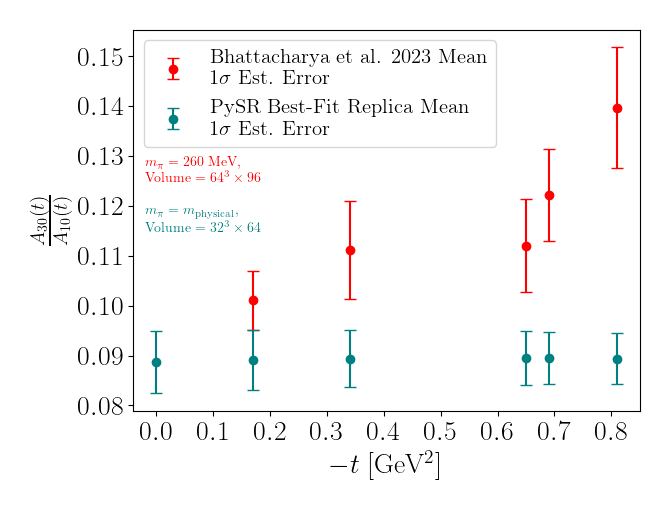}
\caption{{\it Left}: The reconstructed ratio $A_{20}(t)/A_{10}(t)$ from the produced SR results after training and extrapolating $H_{u-d}(x,t)$ in the $x$ region from the reference lattice data set~\cite{Lin:2020rxa} (blue) and comparing with available lattice data on $A_{20}(t)$ and $A_{10}(t)$ (red) Ref.~\cite{Alexandrou:2019ali,Alexandrou:2018sjm}. Both ratios are based on LQCD data obtained at the physical pion mass for a better and accurate comparison;
{\it Right}: Comparison of ratios $A_{30}(t)/A_{10}(t)$ between lattice data from Ref.~\cite{Bhattacharya_2023} computed at higher than the physical pion mass $m_\pi=260$ MeV (red) and the 
636 Best Fit PySR replica with finite $A_{10}(t)$ and $A_{30}(t)$ from data at the physical pion mass (blue). The SR ratios in both plots were generated at the same $-t$ values where the lattice data is available.}
%
\label{fig: moment_ratios}
\end{figure}

An even more stringent test of factorization that captures a broader range of the phase space covered by LQCD, is obtained by inspecting the behavior of the ratios of GPD moments, Eqs.\eqref{eq:mom1} and \eqref{eq:mom2}, for which LQCD results are available ,
\begin{equation}
\label{eq:lattice_ratios}
    \frac{A_{20}(t)}{A_{10}(t)}, \quad \quad \frac{A_{30}(t)}{A_{10}(t)}  .
\end{equation}
Deviations of the ratios from a constant behavior in $| t |$ signal a breaking of factorization.
In Fig.~\ref{fig: moment_ratios}, we show these ratios for moments $A_{i0}(t)$ for $i\in\{1,2,3\}$ provided by LQCD \cite{Alexandrou:2018sjm,Alexandrou:2019ali, Bhattacharya_2023}, and computed from our 597/1000 PySR BF replicas that display finite behavior in $A_{i0}(t)$ for $i\in\{1,2,3\}$.


The SR results, based on lattice data calculated at the physical pion mass, are compared with two available lattice data sets, Fig.~\ref{fig: moment_ratios}. The first corresponds to the ratio 
$A_{20}/A_{10}$, also computed at the physical pion mass, allowing for a direct and accurate comparison.   
The second involves the ratio $A_{30}/A_{10}$ obtained from simulations performed at a heavier pion mass of $m_\pi=260MeV$. In particular, for the $A_{20}/A_{10}$ ratio, the first moment $A_{10}$ was constructed from the electric, $G_E$ and magnetic $G_M$ form factors of the nucleon from the ETMC Collaboration \cite{Alexandrou:2018sjm}, while $A_{20}$ was taken directly from  Ref~\cite{Alexandrou:2019ali}, 
where the errors were propagated twice, once to estimate the first moment $A_{10}$ and the second time to estimate the ratio $A_{20}/A_{10}$. It is important to mention  
that the errors in the ratio of the lattice data were estimated assuming the data from both moments are uncorrelated, which is not strictly true. These estimates are therefore meant as an approximation for a comparison with our PySR results.
Mean values and uncertainties from SR were produced at the same $t$ values of the LQCD points after training and extrapolating $H_{u-d}(x,t)$ with the methodologies detailed in Section \ref{sec:3}.

From  Fig.\ref{fig: moment_ratios} one can see that, even if the differences in the lattice input between Refs.\cite{Bhattacharya_2023} and \cite{Lin:2020rxa} make a fully quantitative comparison challenging, a non-trivial $|t|$-dependence of moment ratios is incompatible with a factorization ansatz for the unintegrated $H_{u-d}$, the factorization breaking being more pronounced at larger values of $t$ in both cases. {Though the BF replicas do not explicitly factorize in $x$ and $t$ symbolically, they do approximately factorize numerically, as indicated by the near constant value of $A_{20}(t)/ A_{10}(t)$. This is because the BF replicas were trained on a range of lattice points that factorized to within $\approx 10\%$ (Fig. \ref{fig: Lin21_ratios}). These replicas were convergent on the training region $0\leq |t|\leq 1$ GeV$^2$, which resulted in the approximately constant point-by-point uncertainties in Fig. \ref{fig: moment_ratios}.}

On the other side, when analyzing PySR results we should keep in mind that, as shown in Table \ref{tab:mse_stats}, we detected no noticeable preference for the BF exemplars over the FF exemplars. This behavior in the $(x,t)$ region is present for the training data as well, and it will be therefore reflected in the replicas behavior.
It is consistent with a framework where the replicas trained on LQCD $x$-dependent results, approximately factorize, namely, even the BF replicas that were not constrained to factorize still approximately numerically factorize. A further confirmation is also found  in the constant behavior  
on average of both $A_{20}(t)/A_{10}(t)$ and $A_{30}(t)/A_{10}(t)$ with $t$. 
This statement is not in tension with current lattice moment results \cite{Bhattacharya_2023}  that imply global factorization breaking due to being integrated over $x$. This could mean that only in regions of $(x,t)$ where $x\leq 0.255$ and $|t|\geq 1 \: \mathrm{GeV}^2$ may factorization breaking be demonstrated more quantitatively. 
In future work, we may impose conditions in the loss function such as requiring that percent deviation of $R$, defined in Section \ref{subsec: test_factorization}, from constant must be within $10\%$ up to $|t|=1 \: \mathrm{GeV}^2$, beyond which replicas are encouraged to violate factorization.


In summary, from the comparison of LQCD moments with PySR, we can conclude that LQCD data allow for a breaking of factorization in the $x,t$ behavior of GPDs that is entirely focused in the low $x$ region. Because the moments are integrated over the full range $0<x<1$, it is possible that even though factorization is broken globally, $H_{u-d}(x,t)$ still approximately factorize in specific regions in $(x,t)$, such as the region trained on in this work, without being in tension with other LQCD moment results.

Our results provide an unbiased comparison of different physics scenarios underlying the GPD description. In particular, we can infer from our analysis of LQCD results, a picture of the low $x$ behavior consistent with Reggeized models such as GGL and GK where the $x$ and $t$ dependence appears entangled in the exponential behavior of the parametric form, {\it e.g.} $x^{\alpha'(1-x)t}$, while it excludes completely factorized forms of the type in Eq.\eqref{eq:factor}.

\subsection{Extracting Physical Densities}
\label{subsec: bT-space}

As discussed in Section \ref{sec:2}, one of the key properties of GPDs is their impact parameter space representation $\rho_{q, g}(x, {\bf{b}}_{T})$, which corresponds to spatial densities of the quarks and gluons inside the nucleon. The factorization of the $x$ and $t$ dependence of GPDs becomes crucial in impact parameter space due to its consequences for the quark RMS transverse radius $\langle {\bf{b}}^{2}_{T} (x) \rangle^{1/2}$, the ``average radii." One can immediately see from Eq.\ref{eq:radius} that if the $x$ and $t$ dependence factorizes, then the $x$ dependence cancels and the radius is constant in $x$. This would be at variance with the 
QCD-based prediction for the radius, which requires that as $x \rightarrow 1$, $\langle {\bf{b}}_{T} (x) \rangle^{1/2} \rightarrow 0$ (for a review of both theory and experiment see Ref.\cite{Dutta:2012ii}).

\begin{figure}[H]
\centering
\includegraphics[width=0.46\linewidth]{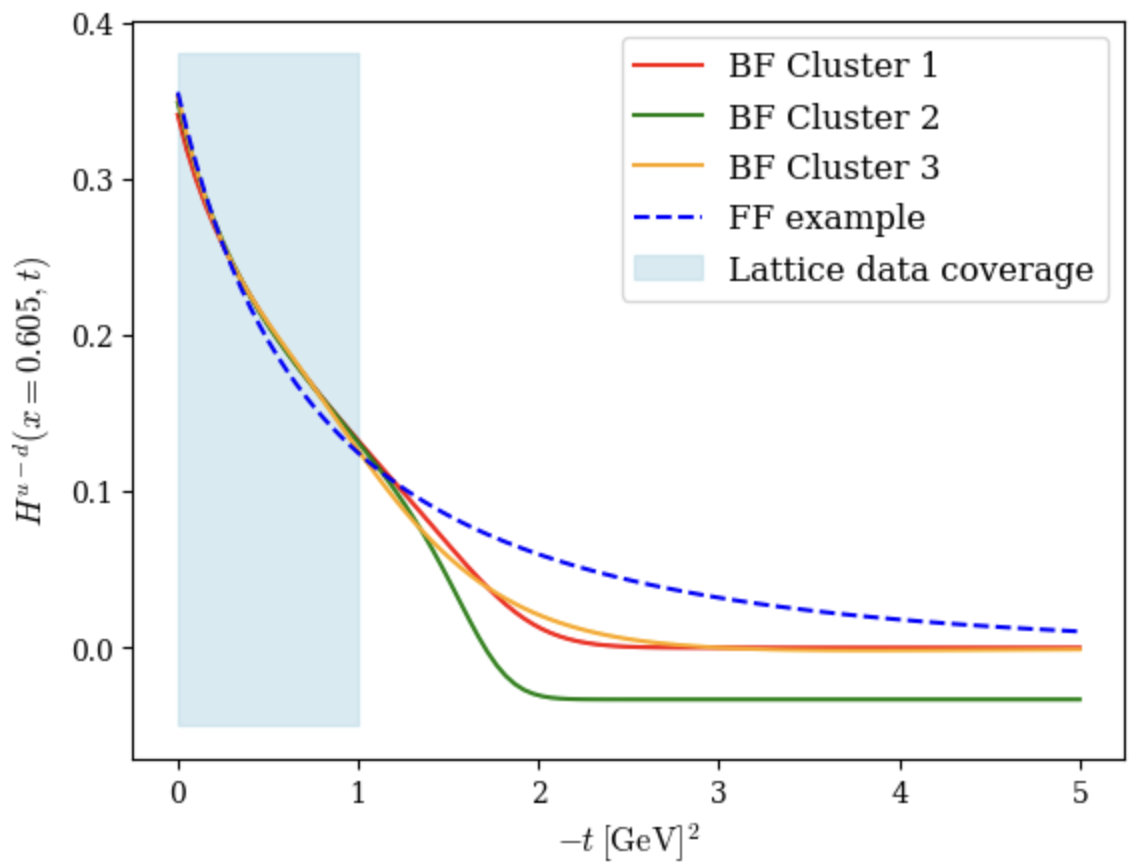}
\includegraphics[width=0.46\linewidth]{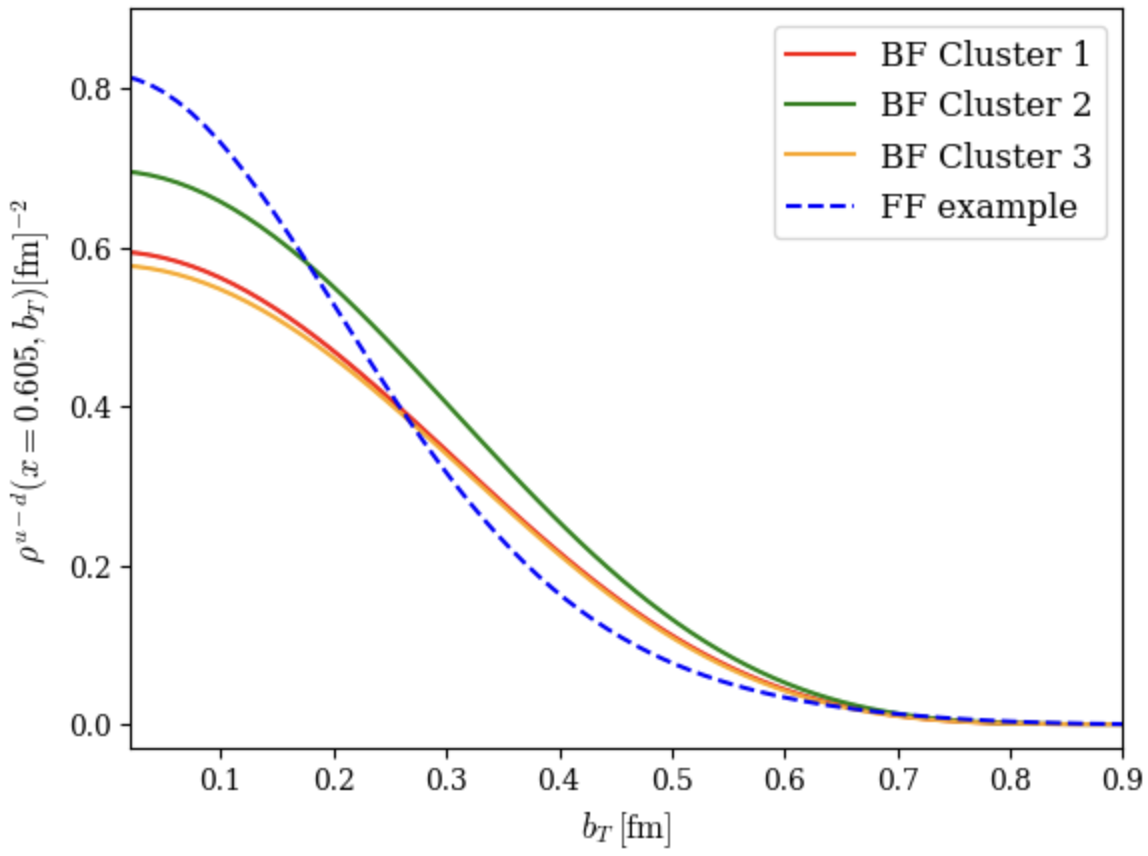}
\caption{Studying the t-dependence and corresponding ${\bf{b}}_{T}$-dependence of the BF models from Table \ref{fig:table} as well as an FF example, Eq. \ref{ff_example}. Although all four distributions match within the region of the training data, they differ in the extrapolated region. That difference propagates in the ${\bf{b}}_{T}$-space and therefore impacts the interpretation of the proton's spatial structure.}
\label{fig: t_bT_dep}
\end{figure}

In Fig.~\ref{fig: t_bT_dep},we present $\rho_{u-d}(x, {\bf{b}}_{T})$, the Fourier transform of the GPDs $H_{u-d}(x, 0, t)$, which is obtained from our SR fit to LQCD results: three representative models obtained through the BF method and clustered by WMSE both within and beyond the range of the data are shown along with a FF model, that we choose as
\begin{eqnarray}
    H_{\mathrm{FF, u-d}}(x, {\bf{t}}) &=& \frac{\Big[ \left( \left(23.9  x+32.5 \right) ^{-1.37} \right)^x \Big]^x - 0.016}{1.67^{\textbf{t}} - 0.638}, 
    \label{ff_example}
\end{eqnarray}
since it predicts a magnitude for the GPD in the extrapolated region similar to the prediction of the BF models. 
From the figure, that juxtaposes the GPD as a function of $|t|$ on the {\it lhs}, to the density on the {\it rhs} for a given value of $x=0.6$,  we can see that GPDs belonging to the range of clusters shown in Figs. \ref{fig:BF_hwl_good_kneedle_spag_b} and 
\ref{fig: FF_hwl_cluster_spag_plot} 
produce the largest differences at small ${\bf b}$, understandably, since this region is governed by large $|t|$, beyond the reach of LQCD results.  

\begin{figure}[H]
\centering
\includegraphics[width=0.5\linewidth]{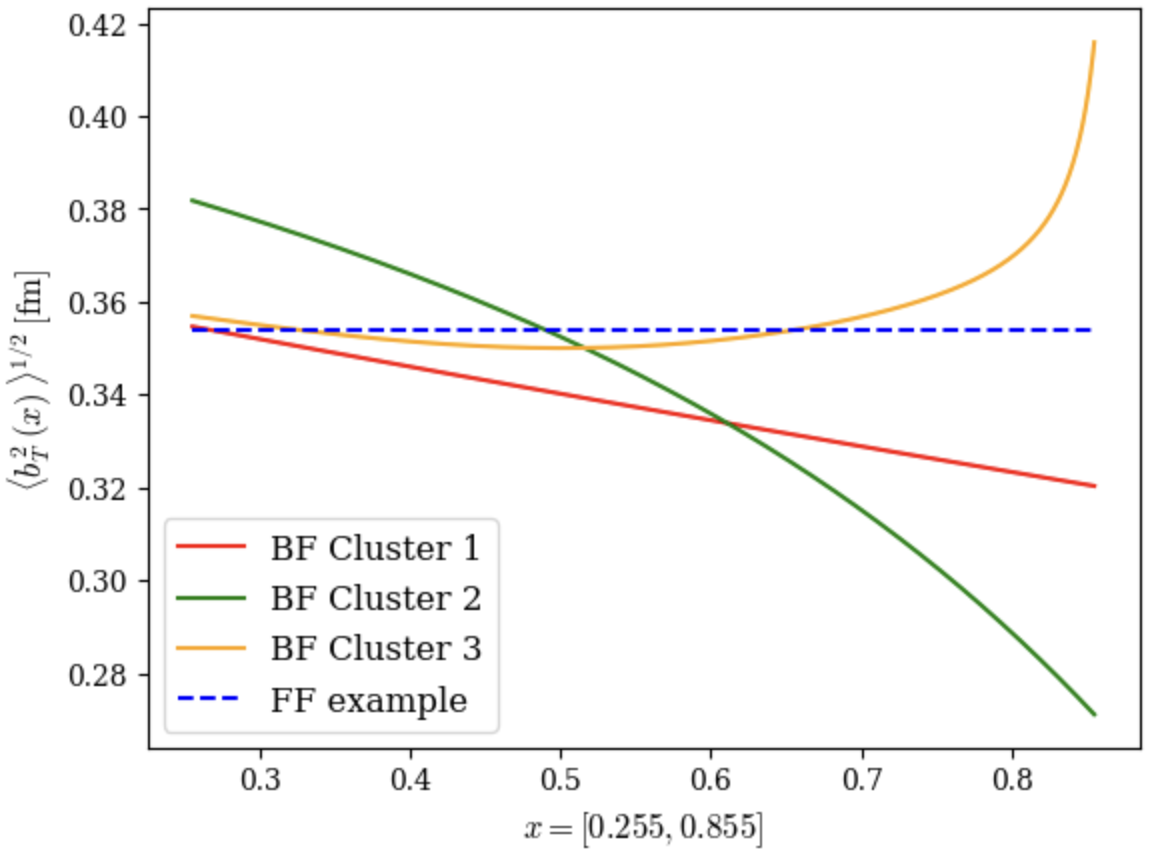}
\caption{The average radii $\langle {\bf{b}}_{T} (x) \rangle^{1/2}$ corresponding to the BF models in Tab. \ref{fig:table} as well as an FF example, Eq. \ref{ff_example}. The expectation from
QCD phenomenology is that as $x \rightarrow 1$,
$\langle {\bf{b}}_{T} (x) \rangle^{1/2}$ → 0.
Notice how the factorized
result is a constant in x,
and BF Cluster 3 turns
upwards as $x \rightarrow 1$. 
We emphasize that all four of these distributions are a prediction of the combined SR \& LQCD framework.}
\label{fig: avg_radii_graph}
\end{figure}

The average radii $\langle {\bf{b}}_{T} (x) \rangle^{1/2}$, Eq.\eqref{eq:radius}, corresponding to the four densities in Fig. \ref{fig: t_bT_dep}, shown in Fig.~\ref{fig: t_bT_dep} and \ref{fig: avg_radii_graph}, illustrate the capability of our SR framework to study different physics scenarios that emerge from the data: the FF one is, of course, constant in $x$, while the curves from the BF WMSE clustering show different trends for the radius of quark and gluon configurations with different momentum fraction, $x$, from fast decreasing as $x\rightarrow 1$, to increasing. The latter would rule out the QCD prediction of point-like configurations \cite{Brodsky:2022bum}. 

\begin{figure}[H]
\centering
\includegraphics[width=0.9\linewidth]{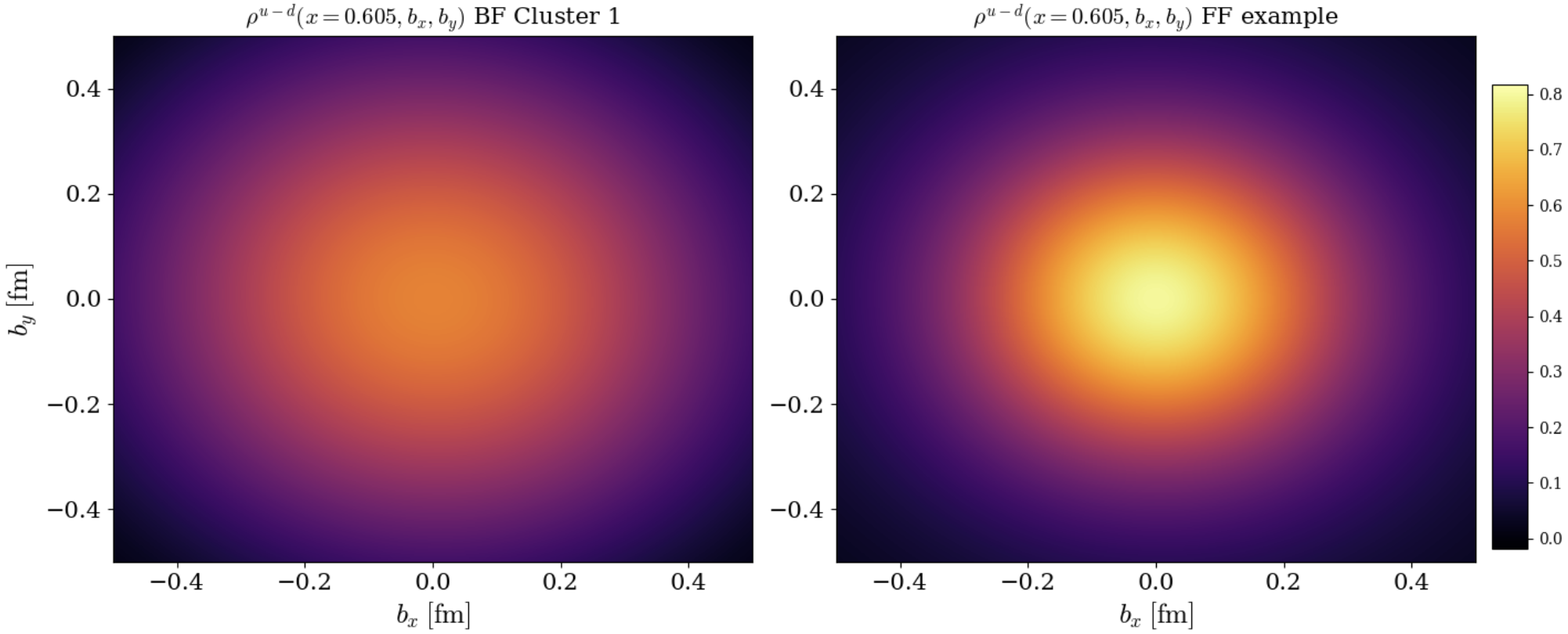}
\caption{The BF Cluster 1 model compared to an example FF model as a function of $b_{x}$ and $b_{y}$. These densities demonstrate that the u and d quarks are more likely to be located at the proton's center of momentum. While these densities offer a useful visualization for investigating the proton's spatial structure, they do not demonstrate fully the implications of having a factorized versus a non-factorized GPD; to do so, we need to explore the average radii.}
\label{fig: 3D_densities}
\end{figure}

\section{Conclusions and Outlook: A roadmap to SR-Assisted Tomography }
\label{sec:conclusions}
We have demonstrated that SR is a powerful new tool that is capable of new tasks that traditional tools could not accomplish (Tab. \ref{fig:table}). 
Our analysis offers new insights on the hadronic physics side where 
we have shown that SR can be fruitfully applied to extract new or better information on the proton 3D quark and gluon substructure than traditional methods (Fig.~\ref{fig: 3D_densities}). Because SR allows for improved extrapolation capabilities (Fig: \ref{fig: t_bT_dep}) beyond the range of LQCD results, we were able to single out a pathway for a quantitative description of  the spatial distributions of quarks and gluons on the transverse plane, extracting, for the first time, quantitative trends of their radii (Fig. \ref{fig: avg_radii_graph}) in different ranges of the momentum fraction, $x$. These trends emerge in the absence of any known priors, using  solely results from LQCD on GPDs and nucleon form factors. 
On the computational, AI/ML side, we have presented a new approach to systematically benchmark, and/or assess the performance of this new tool (Tab. \ref{tab:mse_stats}), including the expansion coefficient clustering (ECC) convergence criterion (Figs. \ref{fig: bestFit_corner_spag_plot},\ref{fig:BF_cluster_spag}, \ref{fig: ForceFactorized_corner_spag_plot}, \ref{fig: FF_hwl_cluster_spag_plot}).

While producing the main results of this work, we have encountered many other instances where our original results can be further elaborated on and/or improved.   
Listing a few in our outlook on future work includes, {\it e.g.}, the implementation of additional physics symmetry constraints in the loss function, and considering the full phase space validation, beyond $t = 0$: as information on the other GPD variables, namely, skewness, $\xi$ and the dependence on the renormalization scale, $\mu^2$ become available, we will also be able to impose proper perturbative QCD evolution in the loss function;  
the SR composite function building blocks can also be further manipulated to understand in more detail the physics underlying the various functional behaviors; our ECC  can be extended to study clustering using other series expansions.

Finally, while community-standard practice generally lacks comprehensive assessment and error analysis of SR techniques, even in one dimension, often providing prediction results without associated error bands,
we are interested in assessing such errors. 
A careful uncertainty quantification investigation deserves an entire new study where we will first study SR on a known fixed dataset, and subsequently consider multi-objective optimization with an associated pareto front analysis. 
%
We will also pursue rigorous uncertainty quantification by comparing to uncertainty quantification in Bayesian neural networks, 
and taking into account the correlations of LQCD results.
%

%
%
This is not yet ``the Answer'' for our extracted GPDs, but this is the tool that will be used to obtain them.
%
%
%
%
%

\acknowledgements
We thank Prasanna Balachandran and Dennis Sivers for fruitful discussions. This work was completed by the EXCLAIM collaboration under the Department of Energy grant DE-SC0024644. We also acknowledge DOE grants DE-SC0016286 (D.A., S.L., Z.P.), DE-SC0024560 (M.S.), PHY 1653405,  PHY 2209424 and  Research  Corporation  for  Science  Advancement through the Cottrell Scholar Award (H.L.), and
SURA Center for Nuclear Femtography (S.L., Y.L., Z.P.). 

\bibliography{references}

\appendix

\section{Charge Conjugation and Isospin Decomposition}
\label{subsec:symm}
The GPD $H^q$ for a given quark flavor $q$ can be combined with the corresponding antiquark distribution $H^{\bar q}$ to form linear combinations which are even ($H_q^+$) or odd ($H_q^-$) under charge conjugation:
\begin{eqnarray}
    H_q^+(x,t)  & = & H_q + H_{\bar{q}},  \quad \quad     H_q^- (x,t)  =  H_q - H_{\bar{q}} \: .
\end{eqnarray}
Note that, since
\begin{align}
    H_{\bar q}(x) =  - H_q(-x)  \: ,
\end{align}
the C-even distributions $H_q^+$ are \textit{anti}symmetric with respect to $x=0$, while the C-odd distributions $H_q^-$ are \textit{symmetric}.

In the forward limit, $(\xi=t=0)$, the quark GPDs $H_q$ match onto the corresponding collinear distributions (see Eq.\ref{eq:pdf}).
As such, the conserved net quantum numbers of the proton $p \simeq (u u d)$ carred by the valence quarks are reflected in QCD sum rules for the PDFs, which therefore constrain the GPDs:
\begin{align}   \label{e:Valence_SumRules}
\begin{aligned}
    \int\limits_0^1 dx \: \Big( u(x, Q^2) - \bar{u}(x, Q^2) \Big) =
    \int\limits_0^1 dx \: H_u^- (x, 0, 0, Q^2) = 2  \: ,
    \\
    \int\limits_0^1 dx \: \Big( d(x, Q^2) - \bar{d}(x, Q^2) \Big) =
    \int\limits_0^1 dx \: H_d^- (x, 0, 0, Q^2) = 1  \: .
\end{aligned}
\end{align}

Writing separately the contribution from quarks/antiquarks that can be generated from gluon splitting and populate the proton sea, $q_s/\bar{q}_s$, and the primordial valence or intrinsic ones, $q_v$,  
 \begin{eqnarray}
 H_q   &  = & H_{q_v} + H_{q_s},  \quad\quad   H_{\bar{q}}  =   H_{\bar{q}_v}+ H_{\bar{q}_s} ,
 \end{eqnarray}
 therefore, taking the Kuti-Weisskopf assumption  that quark-antiquark pairs are generated perturbatively through gluon splitting, therefore, $H_{q_s}= H_{\bar{q}_s}$, one has,  
 \begin{eqnarray}
 H_q^+   &  = & [H_{q_v} + H_{q_s}] +   H_{\bar{q}_s} \equiv  H_{q_v}  +  2 H_{\bar{q}_s}  \\
 H_{q}^- & = &  [H_{q_v} + H_{q_s}]  -  H_{\bar{q}_s}  \equiv H_{q_v}
 \end{eqnarray}
 Analogous expressions can be written for the GPD, $E$. 

We may also decompose the quark flavor dependence into isovector and isoscalar components,
\begin{subequations}
\begin{align}   \label{e:H_Isovector_Defn_app}
    H_{u-d} &\equiv H_u - H_d 
    %
    =
    \frac{H_u^- - H_d^-}{2} + \frac{H_u^+ - H_d^+}{2}
    \: ,
    \\
    H_{u+d} &\equiv H_u + H_d =
    \frac{H_u^- + H_d^-}{2} + \frac{H_u^+ + H_d^+}{2}
    \: ,
\end{align}
\end{subequations}
%

\section{Inputs and Hyperparameters of PySR}
\label{app:Inputs}
%

As discussed in Sec.~\ref{subsec:PySR}, PySR is a multi-population genetic algorithm that generates symbolic mathematical expressions to optimize a loss function with respect to a set of data.  PySR incorporates three unique features into its algorithm: simulated annealing to select the winning expression within an island, an ``Evolve-Simplify-Optimize'' loop when applying the genetic algorithm to a given island, and a strategy of ``adaptive parsimony'' to maintain a good population of expressions at each complexity.

Simulated annealing is a well-established global optimization technique \cite{Pincus:1970sa, Kirkpatrick:1983sa} for avoiding the problem of becoming trapped in a local minimum.  Rather than accepting the expression with the minimized loss as the winner with $p = 100\%$ probability, one takes a Boltzmann form $p = \exp( \Delta l / \alpha T)$ where $\Delta l$ is the change in the expression's loss and $\alpha T$ plays the role of a temperature.  For each mutation step $k = 1 \cdots n_c$, with $n_c = 300,000$ by default, PySR takes the temperature to be $T = 1 - k / n_c$; this simulates the ``cooling'' from high to low temperatures when searching for the expression with the globally minimum loss.  The overall temperature scale of the simulated annealing is set by $\alpha$, which is taken to be $0.1$ by default.

The ``Evolve-Simplify-Optimize'' loop of PySR specializes the classic evolutionary algorithm for use on equations.  This three-step process first takes each expression in an island through $n_c$ steps of random mutations, only thereafter attempting to analytically simplify equivalent mathematical expressions.  Finally, the algorithm optimizes the arbitrary constants in each expression to fit the data.  

Finally, whereas a traditional ``fixed parsimony'' penalty to the loss function (proportional to the complexity of an expression) results in few candidate expressions at higher complexity, PySR chooses a novel ``adaptive parsimony'' approach.  PySR adaptively tunes the per-complexity penalty so that the number of expressions at each complexity is approximately the same.  One can loosely think of this as a ``chemical potential for complexity,'' with the quantity governing the response to the complexity known as the ``frecency.''

\begin{table}[t]
\begin{tabular}{|p{10em}|p{5em}|p{5em}|p{30em}|}
    \hline
    Parameter & Default & This Study & Description   \\
    \hline
    alpha & 1.574 & 1.574 & 
    Initial temperature for simulated annealing (requires annealing to be True).    
    \\  \hline
    annealing & False & False & 
    Whether to use annealing.    
    \\  \hline
    constraints & Not Used & $\{``\text{\^{}}": (-1,1)\}$ & Dictionary of int (unary) or 2-tuples (binary), this enforces maxsize constraints on the individual arguments of operators.\\
    \hline
    binary\_operators & *, /, +, - & *, /, +, -, \text{\^{}}& 
    List of strings for binary operators used in the search.    
    \\  \hline
    crossoverProbability & 7.110e-05 & 7.110e-05 & 
    Absolute probability of crossover-type genetic operation, instead of a mutation
    \\  \hline
    fractionReplaced & 1.337e-02 & 1.337e-02 & 
    How much of population to replace with migrating equations from other populations.
    \\  \hline
    fractionReplacedHof & 3.098e-02 & 3.098e-02 & 
    How much of population to replace with migrating equations from hall of fame.
    \\  \hline
    maxsize & 30 & 50 & 
    Max complexity of an equation.  You can use either maxsize or maxdepth.     
    \\  \hline
    model\_selection & 'accuracy' & 'accuracy' & 
    Model selection criterion when selecting a final expression from the list of best expression at each complexity. Can be 'accuracy', 'best', or 'score'. 
    \\  \hline
    ncyclesperiteration & 2559 & 2559 & 
    Number of total mutations to run, per 10 samples of the population, per iteration.
    \\  \hline
    niterations & 4.0e1 & 1.0e03 & 
    Number of iterations of the algorithm to run. The best equations are printed and migrate between populations at the end of each iteration.
    \\  \hline
    npop & 35 & 35  & 
    Number of individuals in each population.    
    \\  \hline
    optimize\_probability & 1.489e-02 & 1.489e-02 & 
    Probability of optimizing the constants during a single iteration of the evolutionary algorithm.
    \\  \hline
    optimizer\_iterations & 6.0 & 6.0 & 
    Number of iterations that the constants optimizer can take.
    \\  \hline
    optimizer\_nrestarts & 6.0 & 6.0 & 
    Number of time to restart the constants optimization process with different initial conditions.
    \\  \hline
    parsimony & 1.257e-03 & 1.257e-03 & 
    Multiplicative factor for how much to punish complexity.
    \\  \hline
    perturbationFactor & 43.86 & 43.86 & 
    Constants are perturbed by a max factor of $(\mathrm{perturbationFactor}*T + 1)$. Either multiplied by this or divided by this.
    \\  \hline
    populations & 23.0 & 23.0 & 
    Number of populations running.
    \\  \hline
    topn & 32 & 32 & 
    How many top individuals migrate from each population.
    \\  \hline
    tournament\_selection\_p & 0.8375 & 0.8375 & 
    Probability of selecting the best expression in each tournament. The probability will decay as $p*(1-p)^n$ for other expressions, sorted by loss.
    \\  \hline
    unary\_operators & sin, cos, exp, log & Not Used & 
    Operators which only take a single scalar as input.
    \\  \hline
    useFrequency & False & False & 
    Whether to measure the frequency of complexities, and use that instead of parsimony to explore equation space. Will naturally find equations of all complexities.
    \\  \hline
    warmupMaxsizeBy & 3.117e-02 & 3.117e-02 & 
    Whether to slowly increase max size from a small number up to the maxsize (if greater than 0). If greater than 0, says the fraction of training time at which the current maxsize will reach the user-passed maxsize.
    \\  \hline
    weightAddNode & 4.523e-03 & 4.523e-03 & 
    Relative likelihood for mutation to add a node.
    \\  \hline
    weightDeleteNode & 14.48 & 14.48 & 
    Relative likelihood for mutation to delete a node.
    \\  \hline
    weightDoNothing & 1.375e-03 & 1.375e-03 & 
    Relative likelihood for mutation to leave the individual.
    \\  \hline
    weightInsertNode & 27.52 & 27.52 & 
    Relative likelihood for mutation to insert a node.
    \\  \hline
    weightMutateConstant & 3.491 & 3.491 & 
    Relative likelihood for mutation to change the constant slightly in a random direction.
    \\  \hline
    weightMutateOperator & 5.418e-03 & 5.418e-03 & 
    Relative likelihood for mutation to swap an operator.
    \\  \hline
    weightRandomize & 1.629 & 1.629 & 
    Relative likelihood for mutation to completely delete and then randomly generate the equation
    \\  \hline
    weightSimplify & 0.002 & 0.002 & 
    Relative likelihood for mutation to simplify constant parts by evaluation
    \\  \hline
\end{tabular}
\caption{   Hyperparameters, their descriptions, default values in PySR v0.19.4, and the values used in this study.
\label{t:hyperpars}
}
\end{table}

The various inputs and hyperparameters governing PySR v0.19.4 which was used for this study are shown in Tab.~\ref{t:hyperpars} to 4 significant figures.  As recommended in the PySR documentation \cite{SR:2023url}, we have defined the power operator $\mathrm{Pow}[x,y]$ so that base $x$ can have any complexity, but the power $y$ is limited to just complexity $1$.  That is, we only permit expressions with simple powers like $a^x$ or $(2x+1)^y$, but we disallow forms with expressions in the exponent, like $a^{x+y}$.  Note that this does still permit \textit{nesting} of simple powers, such as $((e \wedge x) \wedge x) \wedge x$.

\end{document}